\documentclass[12pt]{iopart}
\usepackage[english]{babel}
\usepackage{iopams}

\expandafter\let\csname equation*\endcsname\relax
\expandafter\let\csname endequation*\endcsname\relax
\usepackage{amsmath}    
\usepackage{amssymb}
\usepackage{amsfonts}
\usepackage{bm}         
\usepackage{bbm}        
\usepackage{dsfont}     
\usepackage{soul}

\usepackage{graphicx}   
\usepackage{xcolor}     
\usepackage{hyperref}   
\usepackage{cite}       

\usepackage[symbol, perpage, hang]{footmisc}

\begin{document}

\title[Ergodicity Breaking and High-Dimensional Chaos in RRN]{Ergodicity Breaking and High-Dimensional Chaos in Random Recurrent Networks}

\author{
Carles Martorell$^{1, \dagger}$, Rub\'en Calvo$^1$, Adri\'an Roig$^1$, Alessia Annibale$^2$, and Miguel A. Mu\~noz$^1$
}

\address{$^1$ Departamento de Electromagnetismo y F\'isica de la Materia and Instituto Carlos I de F\'isica Te\'orica y Computacional, Universidad de Granada, E-18071 Granada, Spain}
\address{$^2$ Department of Mathematics, King's College London, London WC2R 2LS, United Kingdom}
\ead{$^\dagger$carlesma@ugr.es}

\vspace{10pt}
\begin{indented}
\item[] \today
\end{indented}

\begin{abstract}
The neural model introduced by Sompolinsky, Crisanti, and Sommers (SCS) nearly four decades ago has become a paradigmatic framework for studying complex dynamics in random recurrent networks. In its original formulation, with balanced positive and negative couplings, the model exhibits two phases: a quiescent regime, where all activity ceases, and a regime of ongoing irregular collective activity, termed asynchronous chaos (AC), in which state variables fluctuate strongly in time and across units but average to zero across the network. Building on recent work, we analyze an extension of the SCS model that breaks this coupling balance, yielding a richer phase diagram. In addition to the classical quiescent and AC phases, two additional regimes emerge, marked by spontaneous symmetry breaking. In the persistent-activity (PA) phase, each unit settles into a distinct, stable activation state. In the synchronous-chaotic (SC) phase, dynamics remain irregular and chaotic but fluctuate around a nonzero mean, generating sustained long-time autocorrelations. Using analytical techniques based on dynamical mean-field theory, complemented by extensive numerical simulations, we show how structural disorder gives rise to symmetry and ergodicity breaking. Remarkably, the resulting phase diagram closely parallels that of the Sherrington–Kirkpatrick spin-glass model, with the onset of the SC phase coinciding with the transition associated with replica-symmetry breaking. All key features of spin glasses, including ergodicity breaking, have clear counterparts in this recurrent network context, albeit with crucial idiosyncratic differences,  highlighting a unified perspective on complexity in disordered systems.
\end{abstract}

\vspace{2pc}

\submitto{\JSTAT}

\section{Introduction}

The recurrent neural network model introduced by Sompolinsky, Crisanti, and Sommers (SCS) \cite{sompolinsky_chaos_1988} established a theoretical framework for understanding high-dimensional chaos in randomly connected neural networks. In the thermodynamic limit, the model admits an exact dynamical mean-field (DMF) description, in which the combined influence of all neighbors on a representative unit is captured by a single autocorrelated mean field, determined self-consistently by its mean and two-time autocorrelations \cite{sompolinsky_relaxational_1982,sompolinsky_dynamic_1981}. This formulation enables a precise characterization of the system’s phase diagram, including the transition from a quiescent phase—where all dynamics cease—to a chaotic regime with ongoing highly irregular activity.
This seminal result, i.e., the elucidation of high-dimensional chaos, catalyzed a major line of research on self-organized complexity in neural circuits, showing how rich and often unpredictable activity patterns can arise spontaneously from deterministic network dynamics, without finely tuned inputs or control \cite{kadmon_efficient_2025}. In doing so, the SCS forged a conceptual bridge between statistical physics and theoretical neuroscience.

This framework has since been extended to a variety of models with intrinsically generated variability, proving useful for understanding collective computational capabilities in both artificial and biological neural networks. Notable extensions include networks driven by time-dependent inputs \cite{schuecker_optimal_2018, rajan_stimulus-dependent_2010, kadmon_efficient_2025}; 
models with more sophisticated transfer functions
\cite{pazo_discontinuous_2024}; architectures with distinct excitatory and inhibitory populations inspired by biology \cite{van_vreeswijk_chaos_1996, harish_asynchronous_2015, kadmon_transition_2015, mastrogiuseppe_intrinsically-generated_2017}; and networks with more structured synaptic connectivity \cite{mastrogiuseppe_linking_2018, stern_reservoir_2023, wardak_extended_2022, martorell_dynamically_2024, dick_linking_2024, metz_dynamical_2025}.

Here, we analyze an extension of the original SCS model that introduces a non-vanishing average positive coupling between pairs, thereby breaking the balance between excitatory and inhibitory interactions. This setting can be regarded as a specific case of a broader class of network modifications—namely, the addition of low-rank structure to otherwise random connectivity \cite{mastrogiuseppe_linking_2018}. Recent work \cite{mastrogiuseppe_linking_2018, martorell_dynamically_2024} has shown that such structure enriches the phase diagram, generating novel phases absent in the original model, such as one with persistent but non-fluctuating activity, and allows drawing intriguing parallels with the physics of disordered systems.

In this work, we extend previous analysis, studying in detail the emergence of an intermediate phase arising between asynchronous chaos and persistent activity (see Fig.\ref{fig:1}, A). This regime corresponds to a phase of synchronous chaos, where symmetry breaking coexists with chaotic fluctuations. The conceptual origin of this phase can be traced back to the work of Fournier \textit{et al.}, who identified analogous symmetry-broken chaotic states in related mean-field formulations \cite{fournier_non-reciprocal_2025}. While related forms of “structured" or "ferromagnetic" chaos have been reported in \cite{mastrogiuseppe_linking_2018, fournier_non-reciprocal_2025, metz_dynamical_2025}, our analysis provides a more comprehensive characterization, uncovering key dynamical features --such as ergodicity breaking, symmetry breaking, and long-term memory—that, to the best of our knowledge, have not been systematically analyzed within the SCS framework.

We derive closed-form expressions for all the phase boundaries, demonstrating that the resulting phase diagram closely mirrors that of the Sherrington–Kirkpatrick spin-glass model \cite{sherrington_solvable_1975, nishimori_statistical_2001}, and we highlight both striking analogies and fundamental differences between these two paradigmatic statistical-mechanics systems.

The article is structured as follows. Section \ref{sec:dmft_sk} introduces the generalized SCS model and its dynamical mean-field formulation in the thermodynamic limit. Section \ref{Sec. fixed-point analysis} examines the fixed-point structure and establishes the classical phase boundaries. Section \ref{sec:nonstationary} extends the analysis beyond fixed points to characterize the full steady-state dynamics, including the onset of chaos. Section \ref{subsec:theory_SC_main} is devoted to the synchronous-chaotic phase, providing a detailed account of its dynamical signatures: symmetry breaking and ergodicity loss. Section \ref{sec: discussion} presents the conclusions and discusses the broader implications of our findings. Mathematical derivations and additional analyses are provided in the Appendices and Supplemental Material \cite{martorell_supplemental_2025}; for the sake of self-containment and pedagogical clarity, we present these calculations in detail, even when many of them follow standard results already established in the literature.

\section{SCS model: phase diagram and dynamic mean field theory formulation}
\label{sec:dmft_sk}
We define the generalized Sompolinsky–Crisanti–Sommers (SCS) model  \cite{sompolinsky_chaos_1988} by considering a large recurrent network of $N$ neurons. The state of neuron $i \in \{1, \dots, N\}$ is described by a continuous  variable $x_i(t)$, whose dynamics evolves according to the set of differential equations:
\begin{equation}
\partial_t x_i(t) = -x_i(t) + \phi\left( \sum_{j=1}^N W_{ij} x_j(t) \right) 
\label{eq:scs_dynamics}
\end{equation}
where, $\partial_t$ denotes the time derivative, $\phi(x)$ is a nonlinear gain function 
 here taken as $\tanh(x)$,
and $W_{ij}$ are quenched synaptic weights drawn independently from a Gaussian ensemble with mean $\overline{W_{ij}} = J_0/N$ and variance $\overline{W_{ij}^2} - (J_0/N)^2 = J^2/N$,  where $\overline{\; \cdot \;}$ denotes the average over quenched disorder. \footnote{We note that in our formulation the sum over inputs is taken inside the nonlinearity, whereas in the original SCS model the sum was outside. It has been shown, however, that both versions are equivalent leading to the same macroscopic dynamical behavior  in the large-$N$ limit (see  \cite{martorell_dynamically_2024}).}
This scaling ensures that the total input to each unit remains finite in the limit $N \longrightarrow \infty$. In general, $W_{ij} \neq W_{ji}$, so that interactions are non-reciprocal. Extensions of Eq. (\ref{eq:scs_dynamics})
including an additional source of extrinsic stochasticity have also been investigated in \cite{schuecker_optimal_2018, martorell_dynamically_2024}. For clarity, we refer to the state variable as “activity”; however, since it may assume negative values, it should not be interpreted literally as a neuronal firing rate, but rather as a dynamical variable describing the network state.

In the thermodynamic limit $N \longrightarrow \infty$, the dynamics admit an effective decoupling: each unit is governed by a stochastic process driven by a deterministic mean-field component and a stochastic term that together capture the influence of the network. Within the dynamical mean-field (DMF) framework--originally introduced in \cite{sompolinsky_chaos_1988} and subsequently extended by using path-integral methods-- the dynamics of this representative unit are described by the stochastic DMF equation \footnote{For the sake of completeness, a detailed pedagogical derivation is provided in the SM \cite{martorell_supplemental_2025}, where we use a path-integral framework \cite{martin_statistical_1973,de_dominicis_field-theory_1978} (see also \cite{coolen_ch_2001, hatchett_asymmetrically_2004, crisanti_path_2018, helias_statistical_2020, martorell_dynamically_2024
,hertz_path_2017,zou_introduction_2024,galla_generating-functional_2024}).}:
\begin{equation}
\partial_t  x(t) = - x(t) +  \phi \left( g J_0 M(t) + g J \, \eta(t) \right) 
\label{eq:dmft_eq}
\end{equation}
where $\eta(t)$ is a Gaussian process with $\langle \eta(t) \rangle = 0$ and $\langle \eta(t)\eta(s) \rangle = C(t, s)$, and the mean activity $M(t)$ and the autocorrelation $C(t,s)$ are fixed by self-consistency equations:
\begin{equation}
M(t) = \langle x(t) \rangle, \qquad C(t,s) = \langle x(t) x(s) \rangle,
\end{equation}
with averages $\langle \cdot \rangle$ taken over realizations of the effective Gaussian noise, $\eta(t)$. 

\section{Phases and phase transitions: fixed-point  and computational analyses}
\label{Sec. fixed-point analysis}

By assuming that the stationary solution corresponds to a fixed point, $M(t) = M$ and $C(t,s) = q$ for arbitrary times $t$ and $s$, and averaging over the noise, the DMF Eq. (\ref{eq:dmft_eq}) directly yields the following pair of coupled equations for the first two moments:
\begin{align}
M &= \int_{-\infty}^{\infty} D\eta \; \phi\left( g J_0 M + g J \sqrt{q} \, \eta \right), \label{Eq. M}\\
q &= \int_{-\infty}^{\infty} D\eta \; \phi^2\left( g J_0 M + g J \sqrt{q} \, \eta \right), \label{Eq. q}
\end{align}
where $u$ is a Gaussian random variable with zero mean and unit variance, i.e., $D\eta = \frac{e^{-\eta^2/2} \, d \eta}{\sqrt{2\pi}}$ is the standard Gaussian measure. Remarkably, these equations coincide exactly with the replica-symmetric solutions of the Sherrington–Kirkpatrick (SK) spin glass model \cite{sherrington_solvable_1975}, where Eq. (\ref{Eq. M}) corresponds to the magnetization, while Eq. (\ref{Eq. q}) plays the role of the spin-glass order parameter  \cite{edwards_theory_1975, nishimori_statistical_2001}. 

To identify the distinct phases and characterize the transitions between them, we perform a linear stability analysis by Taylor expanding Eqs. \eqref{Eq. M}–\eqref{Eq. q} around the trivial fixed point, corresponding to the quiescent state $M = 0$, $q = 0$ (see \ref{App:SCS_Jacobian}). To leading order, this yields:
\begin{align*}
    M &= g J_0 M + \mathcal{O}(M^2, q^2), \\
    q &= g^2 J^2 q + \mathcal{O}(M^2, q^2),
\end{align*}
from which one readily obtains the limits of stability: 
\begin{itemize}
  \item For \(gJ_0 < 1\) and \(gJ < 1\), the unique stable solution is \(M = q = 0\), which defines the \emph{quiescent} phase, where all dynamics cease.
  
  \item For \(gJ > 1\) with \(gJ_0 < 1\), the autocorrelation $q$ bifurcates to a non-zero value (\(q \neq 0\)) while the mean activity remains zero (\(M = 0\)); this regime corresponds to the \emph{asynchronous-chaos} phase. \footnote{The term \textit{asynchronous} has been widely adopted (see, e.g., \cite{harish_asynchronous_2015,dahmen_second_2019,li_tuning_2020}) to describe the chaotic, weakly correlated dynamical regime emerging in firing-rate models, which reflects the irregular, fluctuation-driven activity observed in cortical networks—commonly referred to as the \textit{asynchronous state} \cite{renart_asynchronous_2010}.}
  
  \item For \(gJ_0 > 1\) with \(gJ > 1\), both the mean and the autocorrelation take non-vanishing values, (\(M,q \neq 0\)), characterizing the \emph{persistent-activity} phase. \footnote{As usual, the mean activity $M$ acts as an order parameter by signaling the spontaneous breaking of the $\pm x$ symmetry, and  both $+M$ and $-M$ constitute admissible solutions of the DMF equations.}
\end{itemize}
\begin{figure}[t]
    \centering
    \includegraphics[width=1 \linewidth]{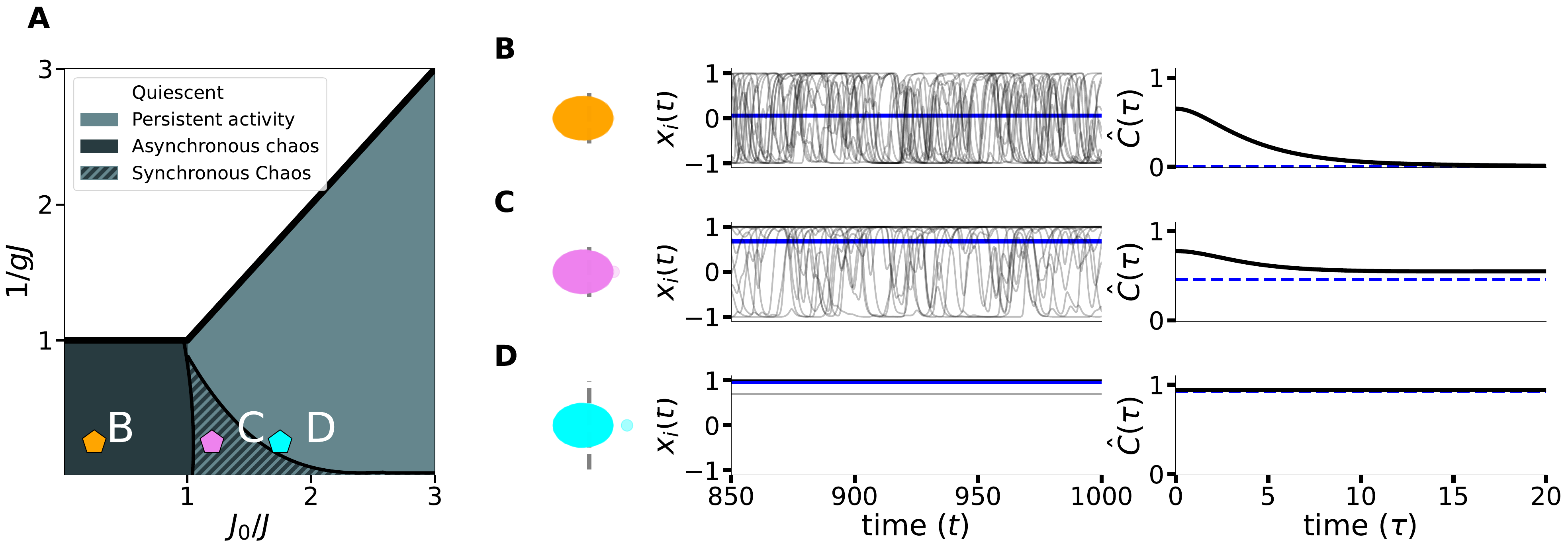}
   \caption{Phase diagram and representative trajectories of the random recurrent network model. 
   \textbf{Panel A}: Phase diagram in the $(J_0 / J, 1/gJ)$ plane, from DMF analyses and confirmed by simulations. Four dynamical regimes are shown with their order parameters: quiescent (Q) $M = 0$, $q = 0$; persistent activity (PA) $M \neq 0$, $q = 0$; asynchronous chaos (AC) $M = 0$, $q \neq 0$, with vanishing long-lag correlations; and synchronous chaos (SC) $M \neq 0$, $q \neq 0$, with persistent correlations. Q–PA and Q–AC transitions follow Eq. \eqref{Eq. fp condition}, while non-quiescent transitions are obtained via steady-state analysis (Sec. \ref{sec:nonstationary}). Markers indicate the parameter choices for panels (B)–(D). \textbf{Panels B/C/D}: 
    Representative states from simulations of the microscopic dynamics, Eq.\eqref{eq:scs_dynamics}, with $N = 2000$ neurons. For each state, we show: the eigenvalue spectrum of the synaptic matrix, which in lowermost case includes an outlier; the time evolution of individual unit dynamics $x_i(t)$ along with the population mean $\hat{M}$ (blue line); and the autocorrelation function $\hat{C}(\tau)$ as a function of lag time $\tau$ along with $\hat{M}^2$ (dashed, blue line).}
    \label{fig:1}
\end{figure}

The stability condition of the quiescent phase can be compactly expressed by
\begin{equation} \label{Eq. fp condition}
g \, \max\left(J_0, J\right) < 1 \quad \implies \quad \max\left(\frac{J_0}{J}, 1\right) < \frac{1}{g J},
\end{equation}
which defines two distinct types of phase transitions, Type I and Type II, depending on which term in the inequality is violated \cite{dahmen_second_2019, li_tuning_2020}. Equivalently, these stability boundaries can be derived from an eigenvalue analysis of the quiescent solution by examining the spectral properties of the random connectivity matrix \cite{martorell_dynamically_2024, mieghem_graph_2010, tao_topics_2012, tao_outliers_2013}, providing a complementary perspective on the onset of instability.
In particular, Type I transitions ($J_0 > J$) correspond to the destabilization of an outlier eigenvalue, whereas Type II ones ($J_0 < J$) occur when eigenvalues at the edge of the bulk spectrum cross the stability boundary (Fig.~\ref{fig:1}).

Fig. \ref{fig:1} summarizes the macroscopic behavior of the SCS model in the thermodynamic limit. In Fig. \ref{fig:1}A, the $(J_0/J, 1/gJ)$ plane is divided into four distinct dynamical phases. The transitions separating the quiescent (Q) state from either the asynchronous chaos (AC) phase ---as already described by SCS--- or the phase with persistent activity (PA) are determined by the fixed-point stability condition in Eq.  \eqref{Eq. fp condition}. Notably, these phases and phase transitions precisely mirror those of the SK model, as predicted by the replica-symmetric ansatz, with the gain $g$ serving as an effective inverse temperature. However, as Fig. \ref{fig:1} also illustrates ---and as we examine in detail below--- this is not the full story: an additional intermediate phase arises between AC and PA, which cannot be identified through the previous linear-stability analysis, as it does not share a boundary with the quiescent phase.

Before proceeding with the analytical approach, and to build intuition and illustrate the properties of all non-trivial phases, Figs. \ref{fig:1}B–D show representative time series of a large finite-size system, obtained from computational simulations of the microscopic dynamics Eq. (\ref{eq:scs_dynamics}).
These plots show the activity of individual neurons along with the (empirical) population mean, 
\begin{equation} \label{Eq. empirical M}
    \hat{M} = \frac 1 N \sum_{i = 1}^N \; \frac 1 T \int_{t_0}^{t_0 + T}  x_i(s) \, ds, 
\end{equation}
and the (empirical) population autocorrelation
\begin{equation} \label{Eq. empirical C}
    \hat{C}(\tau) = \frac 1 N \sum_{i = 1}^N \, \frac 1 T \int_{t_0}^{t_0 + T}  x_i(s)\,  x_i(s + \tau) \, ds, 
\end{equation}
over a finite observation window of length $T = 2000$ time-steps for a network of size $N=2000$.

In the AC phase (Fig. \ref{fig:1}B), neuronal activity exhibits chaotic fluctuations with a vanishing population mean (thick blue line, $\hat M = 0$). The autocorrelation is not constant, but a function that decays to zero. In contrast, the PA phase (Fig. \ref{fig:1}D) is characterized by stable, time-independent activity with a non-zero mean ($\hat M \neq 0$) and a constant, non-vanishing  autocorrelation. Between these two regimes, an intermediate phase, referred to as the \textit{synchronous-chaos} (SC) phase (Fig. \ref{fig:1}C), emerges.
\footnote{We refer to this phase as synchronous chaos, to distinguish it from the asynchronous chaos phase. It has also been described as structured or ferromagnetic chaos\cite{mastrogiuseppe_linking_2018, fournier_non-reciprocal_2025, metz_dynamical_2025} .} 
Here, the single-unit dynamics remain chaotic but develops a non-zero mean. The autocorrelation function decays asymptotically to a finite
value above $\hat{M}^2$, revealing excess long-range temporal correlations. 

In summary, the fixed-point analysis successfully delineates the transitions from the Q phase to both AC and PA in terms of the bifurcations of the autocorrelation $q$ and the mean activity $M$. However, this approach fails to capture the boundaries of the intermediate SC phase, underscoring the need for a full dynamical treatment to characterize the phase diagram.

\section{Phases and phase transitions: Beyond fixed-point analysis}\label{sec:nonstationary}

A stationary, non-fixed-point solution $x(t)$ is defined as a solution of the DMF equation (\ref{eq:dmft_eq}) in which the macroscopic observables, such as $M(t)$ and $C(t,s)$, exhibit time-translational invariance. This implies that the mean activity is constant in time, $M(t) \equiv M$, and the autocorrelation depends only on the time difference $\tau = t - s$, such that $C(t,s) \equiv C(\tau)$ and $C(t,t) \equiv C(0)$. From the DMF equation (\ref{eq:dmft_eq}), one obtains self-consistent equations for the macroscopic observables: the mean activity,
\begin{equation}
M = \int_{-\infty}^{\infty} D\eta \; \phi\Big( g J_0 M + g J \sqrt{C(0)} \, \eta \Big) \equiv \langle \phi (z(t)) \rangle, \label{Eq. M 2}
\end{equation}
where $z(t)= g J_0 M + g J \eta(t)$ 
is the stationary effective input with time-translation–invariant statistics and
$\langle \cdot \rangle$ denotes averaging over the noise, and the autocorrelation function,
\begin{equation} \label{Eq. evolution Delta}
   (\partial_t + 1)(\partial_s + 1) C(t - s)  =   (- \partial_\tau^2 + 1) C(\tau) = 
\langle \phi(z(0)) \, \phi(z(\tau)) \rangle.
\end{equation}
which characterizes the time-lag dependent second moment
$C(t-s)=C(\tau)$ \cite{sompolinsky_chaos_1988, schuecker_optimal_2018, crisanti_path_2018, helias_statistical_2020, martorell_dynamically_2024}.
As already realized by SCS, the dynamics of the autocorrelation function $C(\tau)$ can be interpreted as the energy-conserving motion of a particle in an effective potential $V(C; C_0, M)$:
\begin{equation} \label{Eq. descent-gradient expression}
    \partial^2_\tau C(\tau) = -\partial_C V(C(\tau); C_0, M),
\end{equation}
with
\begin{equation} \label{Eq. potential}
    V(C; C_0, M) = -\frac{1}{2} C^2 + \int_0^C d\Tilde{C} \, \Xi(\Tilde{C}; C_0, M),
\end{equation}
where $\Xi(C(\tau); C(0), M) = \langle \phi(z(0)) \phi(z(\tau)) \rangle$ (see \ref{App: Gaussian properties}). The associated initial conditions are, 
\begin{equation}
    C(0) = C_0, \qquad \partial_\tau C(0) = 0.
\end{equation}
Note that the mean activity $M$ is fixed to its self-consistently determined value from Eq. (\ref{Eq. M 2}), and is therefore treated as an external parameter to determine the  shape of the potential. By contrast, the initial condition $C(0)=C_0$ is not predefined but must be determined dynamically \cite{martorell_dynamically_2024}. In practice, the system selects a steady-state solution of Eq. (\ref{Eq. descent-gradient expression}) characterized by a specific initial condition $C_0^{\text{sel}}$ that guarantees admissibility, i.e. bounded trajectories, together with time-translation invariance and stability, as discussed in what follows.

\subsection{Potential landscape and dynamical state selection}

The stationary autocorrelation $C(\tau)$ is determined by the shape of the effective potential $V(C; C_0, M)$. Since a well-defined autocorrelation must satisfy $|C(\tau)| \leq C(0)$, trajectories remain confined to the interval $C \in [-C_0, C_0]$, allowing for three admissible behaviors: fixed points, oscillations, or asymptotic relaxation to a stationary value (a maximum of the potential).
\footnote{As a consequence of the monotonicity and boundedness properties of the \( \Xi \)-function (see \ref{App: Gaussian properties}), admissible states satisfy $C_0 \leq q$, where $q$ denotes the fixed-point autocorrelation given by Eq.\ (\ref{Eq. q}).}
This framework, originally developed for the asynchronous–chaotic regime ($M=0$) \cite{sompolinsky_chaos_1988, schuecker_optimal_2018, crisanti_path_2018, martorell_dynamically_2024}, is here extended to the synchronous–chaotic case. Importantly, the shape of the potential is primarily controlled by $M$, which determines whether it is symmetric or asymmetric.

  \begin{figure}[t]
    \centering
    \includegraphics[width=1 \linewidth]{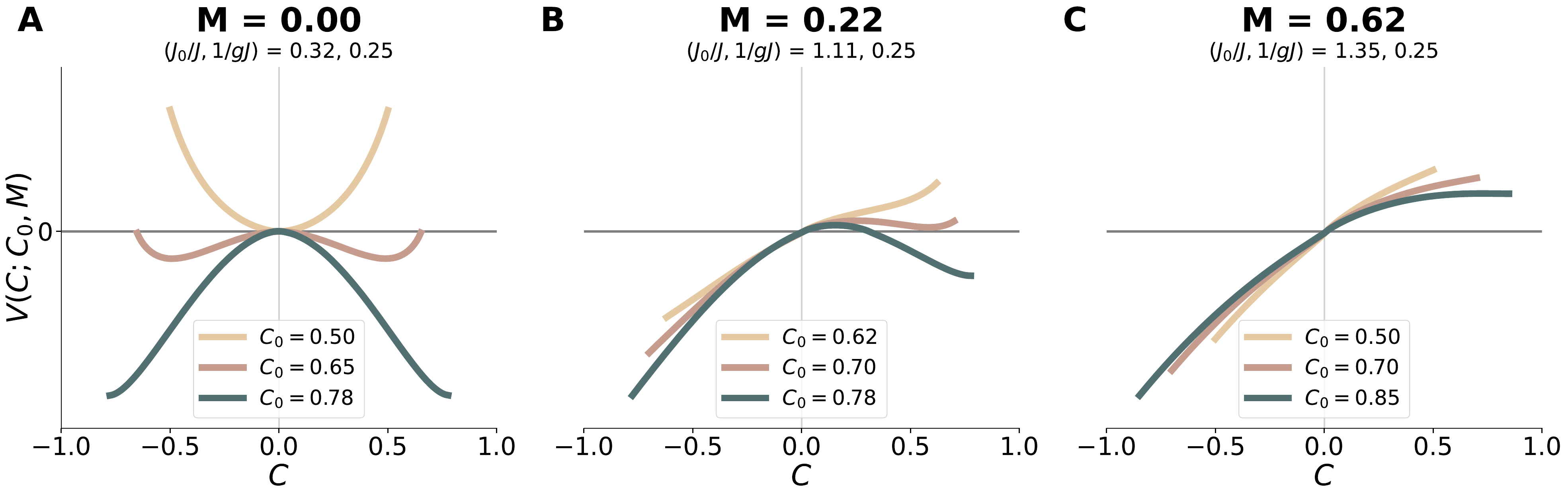}
\caption{Effective potential $V(C; C_0, M)$ as a function of the autocorrelation $C$, shown for different initial conditions $C_0$ (beige, pink, and blue-gray curves) and for representative parameter sets $(J_0/J, 1/gJ)$ corresponding to the AC (panel A), SC (panel B), and PA (panel C) states. In each panel, the order parameter $M$ is fixed to the value dynamically selected by the system in the corresponding state, as determined by Eqs. (\ref{Eq. V'}--\ref{Eq. V''}). The three choices of $C_0$ illustrate key dynamical scenarios: the beige curve corresponds to possible oscillatory (unstable) solutions, the pink curve corresponds to the dynamically selected $C_0$ (in panels A and B) giving a stable solution, and the blue-gray curve shows the fixed-point (at $C=q=0$) value determined by Eq. (\ref{Eq. q}).}
    \label{fig:2}
\end{figure}

Representative shapes of the effective potential are shown in Fig. \ref{fig:2} for increasing values of the mean activity $M$. In the AC regime ($M=0$, panel A), the potential is symmetric with a local extreme (maximum/minimum) at the origin, so that trajectories may oscillate (beige line, $C_0 = 0.50$), remain fixed (blue-gray line, $C_0 = 0.78$), or asymptotically converge to the maximum depending (pink line, $C_0 = 0.65$) on the initial condition. When $M$ bifurcates (SC phase, panel B), the potential becomes asymmetric and develops a displaced maximum at positive values; in this case, trajectories can remain fixed (blue-gray line, $C_0 = 0.78$), or asymptotically converge to the new maximum (pink line, $C_0 = 0.70$). For sufficiently large $M$ (PA phase, panel C), the potential becomes strictly monotonic and admits only a fixed-point trajectory (blue-gray line, $C_0 = 0.85$). 

The stability of the admissible steady states can be determined via linear stability analysis around the stationary autocorrelations, yielding an equation for the Lyapunov exponent (see \ref{App:SCS_Jacobian}). This analysis shows that oscillatory solutions are always unstable, while fixed points are stable only when they constitute the unique bounded solution, as in the PA phase. Consequently, in both the AC and SC phases, oscillatory and fixed-point solutions are unstable, and the dynamics converge to a chaotic state. 
This state is characterized by an autocorrelation function $C(\tau)$ that asymptotically approaches a constant value $C_\infty$, corresponding to a maximum of the potential --a hallmark of chaotic dynamics, as originally established by SCS \cite{sompolinsky_chaos_1988} and further elaborated in \cite{schuecker_optimal_2018, martorell_dynamically_2024}.

Thus, the dynamically selected state in the AC and SC phases corresponds to such a trajectory, uniquely determined by the initial condition $C_0^{\text{sel}}$ and the self-consistently determined value of $M$. The plateau, $C_\infty$, coincides with a local maximum of the potential, defined geometrically through  
\begin{align}
  \partial_C V (C_\infty;C_0^{\text{sel}},M)&=0
     &&\Longleftrightarrow& \Xi(C_\infty;C_0^{\text{sel}},M)&=C_\infty,
     \label{Eq. V'}\\
  \partial_C^2 V(C_\infty;C_0^{\text{sel}},M)&<0
     &&\Longleftrightarrow& \partial_C \Xi(C_\infty;C_0^{\text{sel}},M)&<1,
     \label{Eq. V''}
\end{align}
with the admissible $C_0^{\text{sel}}$ fixed by the energy-conservation condition  
\begin{equation} V(C_0^{\text{sel}};C_0^{\text{sel}},M) = V(C_\infty;C_0^{\text{sel}},M). \label{Eq. energy conservation}
\end{equation}
  
\begin{figure}[t]
    \centering
    \includegraphics[width=\linewidth]{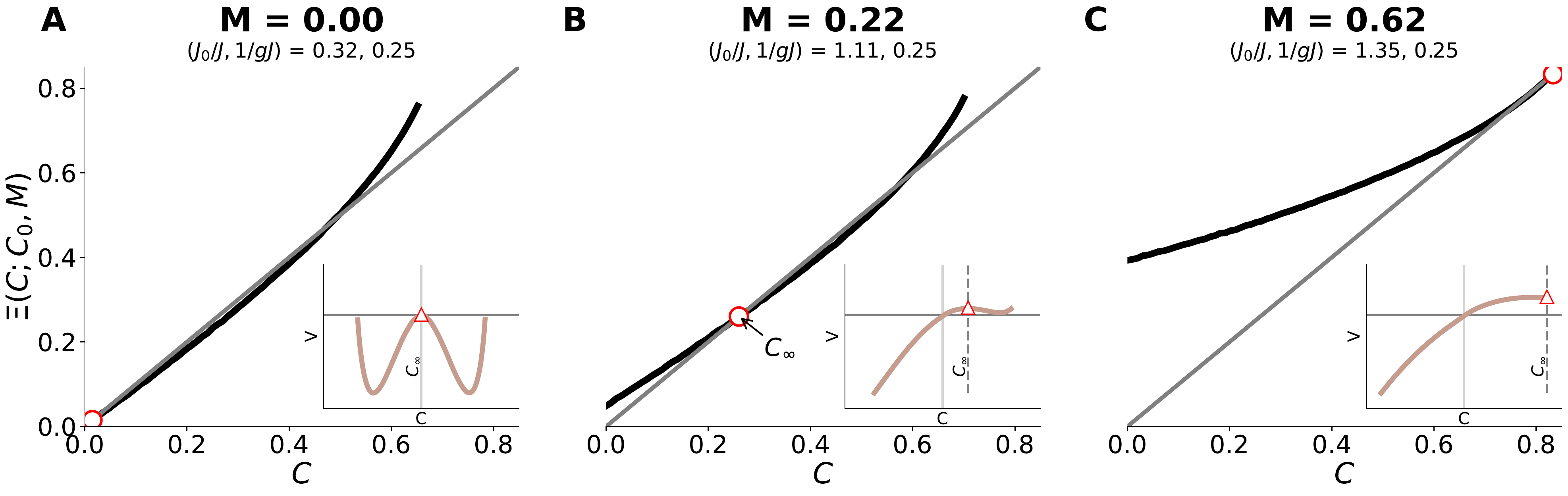}
    \caption{
      Self–consistent evaluation of $\Xi(C;C_0^{\text{sel}},M)$ for the
      dynamically selected values of $C_0^{\text{sel}}$ and $M$ in each phase:
      asynchronous chaos (AC, panel A), synchronous chaos
      (SC, panel B), and persistent activity (PA, panel C).
      Each panel shows $\Xi(C;C_0^{\text{sel}},M)$ (black) together with the
      diagonal $C$ (grey), restricted to $C\in[0,1]$.
      The red circle marks the self–consistent intersection
      $C_\infty$ satisfying Eq. (\ref{Eq. V'}); this point is a
      local maximum of the potential.
      Insets display the corresponding effective potential
      $V(C;C_0^{\text{sel}},M)$ computed at the same parameters; the red
      triangle indicates the local maximum at $C_\infty$.}
    \label{fig:3}
\end{figure}

In summary, the asymptotic state is determined entirely by the geometry of the effective potential. To make explicit the condition in Eq.~(\ref{Eq. V'}), we recast the problem as a graphical rule: for fixed $M$ and $C^{\text{sel}}_0$, the long-time autocorrelation value $C_\infty$ is given by the first intersection of $\Xi(C;C_0,M)$ with the diagonal $x=y$, satisfying $\partial_C \Xi(C;C_0,M)<1$. Figure~\ref{fig:3} illustrates this construction for representative points $(J_0/J,\,1/gJ)$ located at each dynamical phase, where, for clarity, $M$ and $C^{\text{sel}}_0$ are fixed to the dynamically selected values (that can be obtained either computer simulations or solving numerically the equations).

\begin{itemize}
  \item \textbf{AC phase} (\,$M=0$, Fig. \ref{fig:3}A\,):
        $\Xi(C;C_0^{\text{sel}},0)$ intersects the diagonal only at the
        origin, with slope $\partial_C \Xi(0;C_0^{\text{sel}},0)<1$. Thus, $C_\infty = 0$ is the unique stable solution.  

  \item \textbf{SC phase} (\,$M>0$, Fig. \ref{fig:3}B\,):
        because $\Xi(0;C_0^{\text{sel}},M)=M^{2}$, the curve is shifted upward and bends, generating a new crossing at
        $C_\infty>0$ with $\partial_C \Xi(C_\infty;C_0^{\text{sel}},M)<1$. Specifically, the autocorrelation saturates
        at $C_\infty>M^{2}$.

  \item \textbf{PA phase} (large $M$, Fig. \ref{fig:3}C):
        the whole $\Xi(C;C_0^{\text{sel}},M)$ curve falls above the diagonal,
        $\partial_C \Xi(q;q,M)<1$, so the unique intersection is the stable
        fixed point $C_0^{\text{sel}} = C_\infty  \equiv q$. 
\end{itemize}

The geometrical analysis determines the dynamical state selected by the system --in the thermodynamic limit-- and emphasizes the mean activity $M$ as the key control parameter: with $M=0$ the system is chaotic with $C_\infty=0$ (AC); once $M$ bifurcates it remains chaotic but with a finite plateau $C_\infty>0$ (SC); for sufficiently large $M$ the dynamics converge to the fixed‐point PA phase.  In the next subsection we explicitly derive the  phase-transition lines that separate the different non-quiescent phases.

\subsection{Stability criteria determining the SC boundaries}
The SC phase occupies the region between the asynchronous chaos AC and persistent activity PA phases. Its boundary with the PA phase is set by the instability of the PA fixed point, which occurs through a destabilization of the constant equal-time autocorrelation $q$. Conversely, the boundary with the AC phase arises when the stationary solution $M=0$ associated to the AC phase becomes unstable with respect to small fluctuations in the mean activity $M$.

\paragraph{PA\,$\to$\,SC transition.} At the PA phase, the fixed-point solution is defined by
Eqs. \eqref{Eq. M}–\eqref{Eq. q}.  
Linearising the equations around the fixed-point state and retaining the first non-trivial term yields the instability condition (see \ref{App: jacobian_M})
\begin{equation}
   \frac{1}{gJ}
   =\Bigl[\int_{-\infty}^{\infty}\!D\eta\; 
\phi'\!\bigl(gJ_{0}M+gJ\sqrt{q}\,\eta\bigr)^{2}\Bigr]^{1/2},
   \label{eq:PA_SC_main}
\end{equation}
where $\phi'(x)$ denotes the derivative of $\phi(x)$, with $(M,q)$ evaluated at the PA solution. The condition of Eq. \eqref{eq:PA_SC_main}, shown in Fig. \ref{fig:1} as the solid black curve bounding the SC region within the PA phase, was first derived in \cite{mastrogiuseppe_linking_2018} and coincides exactly with the de Almeida–Thouless (AT) line of the SK model \cite{almeida_stability_1978}, marking the onset of replica-symmetry breaking.

\paragraph{AC\,$\to$\,SC transition.}
In the AC phase the mean activity vanishes, \(M=0\), while the zero-lag correlator is fixed at \(C(0) = C_{0}^{\text{sel}}\) --defined by Eqs. (\ref{Eq. V'})-(\ref{Eq. V''})-(\ref{Eq. energy conservation}). A first-order expansion of the mean-activity equation then yields (see \ref{App. stability correlation})
\begin{equation}
   \frac{1}{gJ_{0}}
   \;=\;  \int_{-\infty}^{\infty}\!D\eta\;\phi'\!\bigl(gJ\sqrt{C_{0}^{\text{sel}}}\,\eta\bigr),
   \label{eq:AC_SC_main}
\end{equation}
which marks the symmetry-breaking bifurcation of $M$ that drives the system from the AC state into the SC phase. An equivalent condition was obtained in \cite{mastrogiuseppe_linking_2018} using a different stability approach, as detailed in \ref{App: extended Jacobian analysis}. The corresponding boundary is shown as a solid black line in Fig. \ref{fig:1}A. Thus, the AC phase loses stability when the mean activity $M$ bifurcates to a non-zero value, and concomitantly, the autocorrelation attains a non-zero long-lag value, $C_\infty > 0$.
\section{Synchronous Chaos and Ergodicity Breaking}\label{subsec:theory_SC_main}

In order to characterize the dynamical nature of the SC state, we first focus on its excess of long-time correlations, as anticipated in the previous sections.  Recall that the value of $C_\infty$ is fixed by the geometric constraints of the effective potential: specifically, Eq.\ (\ref{Eq. V'}) determines its value self-consistently, while Eqs.\ (\ref{Eq. V''})–(\ref{Eq. energy conservation})  establish the corresponding stability and energy-conservation conditions.

While in the AC phase ($M = 0$) the self-consistency condition Eq. (\ref{Eq. V'}) admits a unique stable solution with $C_\infty = 0$, in the SC phase—beyond, i.e. the AC–SC transition—the bifurcation of the mean activity to $M \neq 0$ gives rise to a new stable solution that satisfies $C_\infty > M^2 > 0$.  This motivates the definition of an excess-correlation parameter
\begin{equation} \label{Eq. delta}
\Delta = C_\infty - M^2.
\end{equation}
Substituting Eq. (\ref{Eq. delta}) into Eq. (\ref{Eq. V'}) yields a closed self-consistent equation for $\Delta$ within the DMF framework,
\begin{equation}
\Delta = \Xi\bigl(M^2 + \Delta; C_0^\text{sel}, M\bigr) - M^2,
\label{eq:Delta_SC_self}
\end{equation}
the solution of which confirms that, $\Delta$ is strictly positive in the broken symmetry (SC and PA) phases. Thus, $\Delta$ serves as an order parameter to discriminate chaotic phases: $\Delta = 0$ in the AC phase, while $\Delta > 0$ in the SC and PA phase.  

This naturally raises the question of the dynamical origin of the excess correlation $\Delta > 0$,  i.e. why the long-lag autocorrelation saturates at a value larger than $M^2$ in the SC phase.  In the following sections we address this question at the microscopic scale.

\subsection{Ergodicity breaking at the DMF level}
\label{app:decomposition}

In the following, we provide a formal characterization of ergodicity breaking within the DMF framework, following the strategy developed for spin glasses \cite{sompolinsky_relaxational_1982,sompolinsky_dynamic_1981,crisanti_dynamics_1988,crisanti_dynamics_1987}. The analysis relies on the fact that the effective unit $x(t)$ evolves as a stationary, continuous-time Markov process, which allows ergodicity breaking to be rigorously defined from the perspective of stochastic-process theory. The corresponding mathematical formulation is summarized in \ref{app:ergodicity} for completeness and consistency.

Specifically, in the SC phase ($\Delta = C_\infty - M^2 > 0$), we decompose the effective noise into two contributions: a relaxational component, which captures the decay dynamics, and a static component, which accounts for the long-time plateau,
\begin{equation}
\eta(t) = u(t) + \zeta ,
\end{equation}
where $u(t)$  is a zero-mean, stationary Gaussian process with autocorrelation $ \langle u(t)u(t+\tau)\rangle = C(\tau) - C_\infty $ , and $ \zeta \sim \mathcal{N}(0, C_\infty)$. In this representation, $u(t)$ becomes asymptotically uncorrelated, while the variance of the static, node- and realization-dependent term $\zeta$ equals the long-time autocorrelation. This naturally leads to a corresponding decomposition of the effective input current, 
\begin{equation}
   z(t)\;=\;gJ_{0}M \;+\; gJ\,u(t)\;+\; gJ\,\zeta ,
   \label{eq:z_split_main}
\end{equation}
On one hand, $u(t)$ is a stochastic process that decorrelates at long-lag times, capturing temporally correlated fluctuations generated by recurrent interactions; thus, is ergodic [because of the sufficient condition stated by Eq. (\ref{eq:app_int_corr})]. On the other hand, $\zeta$ is a quenched, field arising from the projection of the population mean activity onto each specific neuron through the disordered connectivity. Thus, the DMF equation describes the dynamics of a neuron conditioned on its particular $\zeta$, which reflects both the network realization and the neuron-specific connectivity. In the following, we denote by $\langle \cdot \rangle_u$ and $\langle \cdot \rangle_\zeta$ the average over $u(t)$ and $\zeta$ realizations, respectively. 

Averaging the effective input current, $z(t)$,  over the dynamical component \(u(t)\), while conditioning on a fixed realization of the static variable \(\zeta\), yields a sample-dependent mean,
\begin{equation} \label{eq. mu}
    \mu(\zeta) = g J_0 M + g J \zeta = g J_0 \left( M + \frac{J}{J_0} \zeta \right).  
\end{equation}
This expression highlights that \(\mu(\zeta)\) is a realization-specific shift of the global mean activity \(gJ_0 \, M\), where the perturbation becomes negligible in the limit
\(J_0/J\to \infty\).  The statistics of \(\mu(\zeta)\) over 
disorder realizations are characterized by  
\begin{equation} \label{Eq. statistics mu}
    \langle \mu \rangle_\zeta = gJ_0 M, \qquad \operatorname{Var}_\zeta[\mu] = (gJ)^2 \;  C_\infty = (gJ)^2 \;  (M^2 + \Delta).  
\end{equation}

Consequently, the distribution of $\mu(\zeta)$ across different disorder samples concentrates around \(gJ_0 \, M\) but exhibits a finite width governed by the excess-correlation parameter \(\Delta >0\).

The long-time average of the effective input current is given by
\begin{equation}
\overline{z}_{T}=
\frac{1}{T}\int_{t_0}^{t_0 + T} z(t)\, dt 
\xrightarrow{T \to \infty}
gJ_{0}M + gJ\,\zeta = \mu(\zeta),
\end{equation}
since the long-time average of the dynamical component $u(t)$ vanishes by ergodicity (due to the Birkhoff-Khinchin's theorem, see \ref{app:ergodicity}). Therefore, the long-time average of $z(t)$ does not coincide with its unconditional ensemble average, $\langle z(t)\rangle = gJ_{0}M$, which is independent of the quenched variable. Thus, $z(t)$ is \emph{non-ergodic} with respect to disorder realizations, as time averages retain sample-specific components and fail to self-average in the long-time limit.

A similar analysis applies to the effective neural activity \(x(t)\). For a fixed realization of \(\zeta\), the inner mean is defined as the expectation of the nonlinear output,
\begin{equation} \label{eq: inner mean}
   m(\zeta) = \left\langle \phi\bigl(z(t)\bigr) \right\rangle_{u|\zeta},
\end{equation}
where the average is taken over realizations of the fast component \(u(t)\), conditioned on \(\zeta\). By the Birkhoff-Khinchin's theorem \cite{gardiner_handbook_2004, borovkov_probability_2013}, the long-time average leads to $\overline{\phi(z)}_{T} \xrightarrow{T \to \infty}  m(\zeta)$.
Performing the long-time average on $(\partial_t+1)x(t)=\phi\bigl(z(t)\bigr)$, we obtain 
\begin{equation}
   \overline{x}_{T} \equiv \frac{1}{T} \int_{t_0}^{t_0 + T} x(t)\,dt
   \xrightarrow{T \to \infty} m(\zeta),
\end{equation}
indicating that the long-time average of the effective activity converges almost surely to the sample-dependent inner mean $m(\zeta)$. Averaging over the quenched-disorder distribution (see \ref{app:disorder} for technical details) then yields
\begin{equation} \label{Eq. statistics inner mean}
       M = \langle m(\zeta) \rangle_\zeta, \qquad
   \Delta = \operatorname{Var}_\zeta \bigl[ m(\zeta) \bigr].
\end{equation}
Eqs.~(\ref{Eq. statistics inner mean}) establish the relation between the macroscopic observables, $M$ and $\Delta$, and the microscopic quantity $m(\zeta)$. Different realizations of the quenched disorder---through varying values of $\zeta$--- yield sample-specific mean activities $m(\zeta)$, giving rise to a finite variance $\Delta > 0$. Consequently, the process $x(t)$ is \textit{non-ergodic}: its long-time average depends on the specific realization of $\zeta$ and therefore differs from the unconditional ensemble mean, $\langle x(t) \rangle = M$. The parameter $\Delta$ thus quantifies the degree of ergodicity breaking, measuring the deviation between time and ensemble averages.

In summary, the DMF analysis establishes that the emergence of a non-vanishing excess correlation $\Delta$ and the broad distributions of both inner means, $m(\zeta)$, and mean inputs, $\mu(\zeta)$, are direct consequences of ergodicity breaking. Once $M$ bifurcates, the effective unit acquires a realization-specific inner mean, determined by the quenched variable $\zeta$, such that time averages deviate from disorder averages.

\subsection{Numerical evidence of ergodicity breaking}
\label{ssec:SC_numerical}

\begin{figure}[tbh]
    \centering
    \includegraphics[width=1\linewidth]{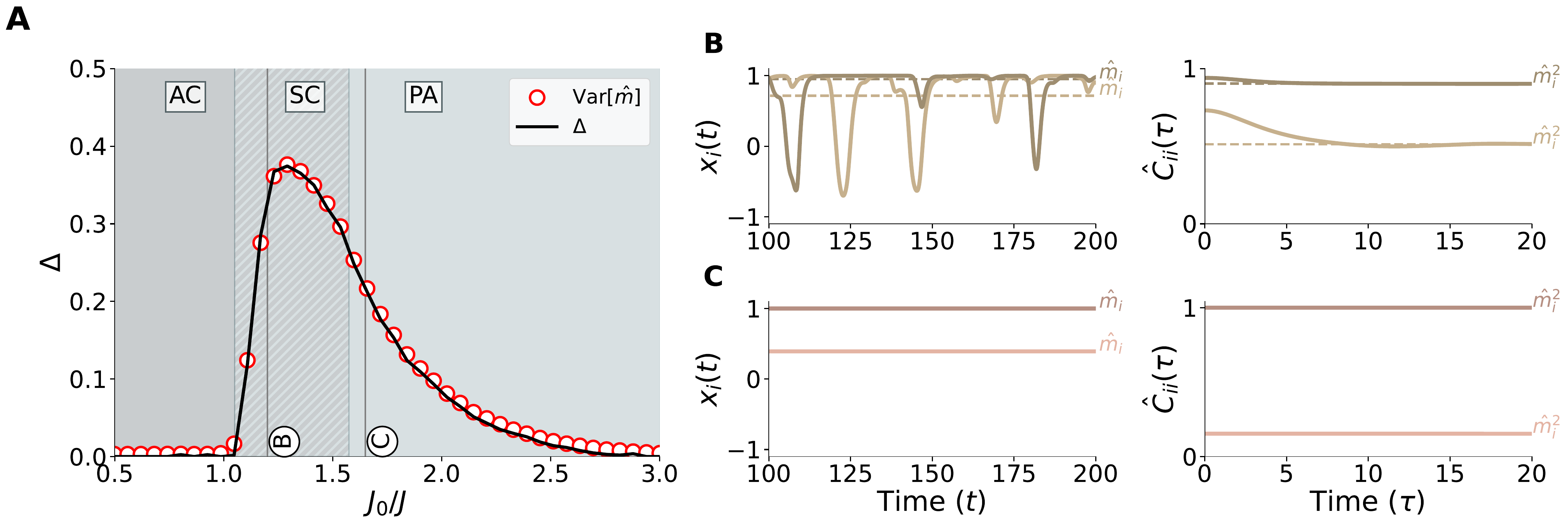}
    \caption{Ergodicity breaking in the broken-symmetry (SC and PA) phases. \textbf{Panel A} shows disorder-averaged statistics of the excess-correlation parameter $\hat{\Delta} = \mathrm{Var}[\hat{m}]$ for $1/gJ = 0.25$ (red circles) together with the analytical prediction $\Delta$ (solid black line) obtained by numerically solving Eq.~(\ref{eq:Delta_SC_self}). \textbf{Panels B–C} display representative trajectories for a single disorder realization at $J_0/J = 1.20$ (SC) and $J_0/J = 1.60$ (PA), showing the activity $x_i(t)$ of two units and their inner means $\hat{m}_i$ (dashed lines). The corresponding autocorrelations $\hat{C}_{ii}(\tau)$ converge to $\hat{m}_i^2$, forming sample-dependent plateaus consistent with $\Delta > 0$. Simulations used $N = 2000$ neurons, $S = 100$ disorder realizations, and trajectory length $T = 2000$ time units.}
    \label{fig:4}
\end{figure}

To validate the DMF predictions, we performed direct simulations of the microscopic dynamics, Eq.~(\ref{eq:scs_dynamics}). For each realization of disorder, the empirical inner mean of neuron $i$ is defined as
\begin{equation} \label{Eq. empirical inner mean}
    \hat{m}_i = \frac{1}{T} \int_{t_0}^{t_0 + T} x_i(s)\, ds,
\end{equation}
and the corresponding empirical excess correlation as
\begin{equation} \label{Eq. empirical Delta}
    \hat{\Delta} = \mathrm{Var}[\hat{m}_i],
\end{equation}
where averages are performed over nodes and disorder realizations. 

Figure~\ref{fig:4}A shows  the empirical excess correlation $\hat{\Delta}$ (red circles) as a function of $J_0/J$ for $1/gJ = 0.25$, compared with the analytical prediction (solid black line) of Eq.~(\ref{eq:Delta_SC_self}). The numerical results reproduce the DMF prediction, vanishing in the AC phase and bifurcating to positive values at the AC--SC transition. 

To illustrate the microscopic origin of this non-ergodic behavior, Fig.~\ref{fig:4}B displays the activity $x_i(t)$ of two representative neurons from a single realization at $(J_0/J,\,1/gJ) = (1.20,\,0.25)$ (SC phase), together with their inner means (dashed lines). The inner means differ across neurons, indicating that long-time averages retain unit-specific components. The corresponding single-unit autocorrelations $\hat{C}_{ii}(\tau)$ also vary across neurons and asymptotically converge to $\hat{m}_i^2$. This heterogeneity persists in the PA phase, as shown in Fig.~\ref{fig:4}C, where each neuron converges to a distinct fixed point.

\begin{figure}[t]
    \centering
\includegraphics[width=1\linewidth]{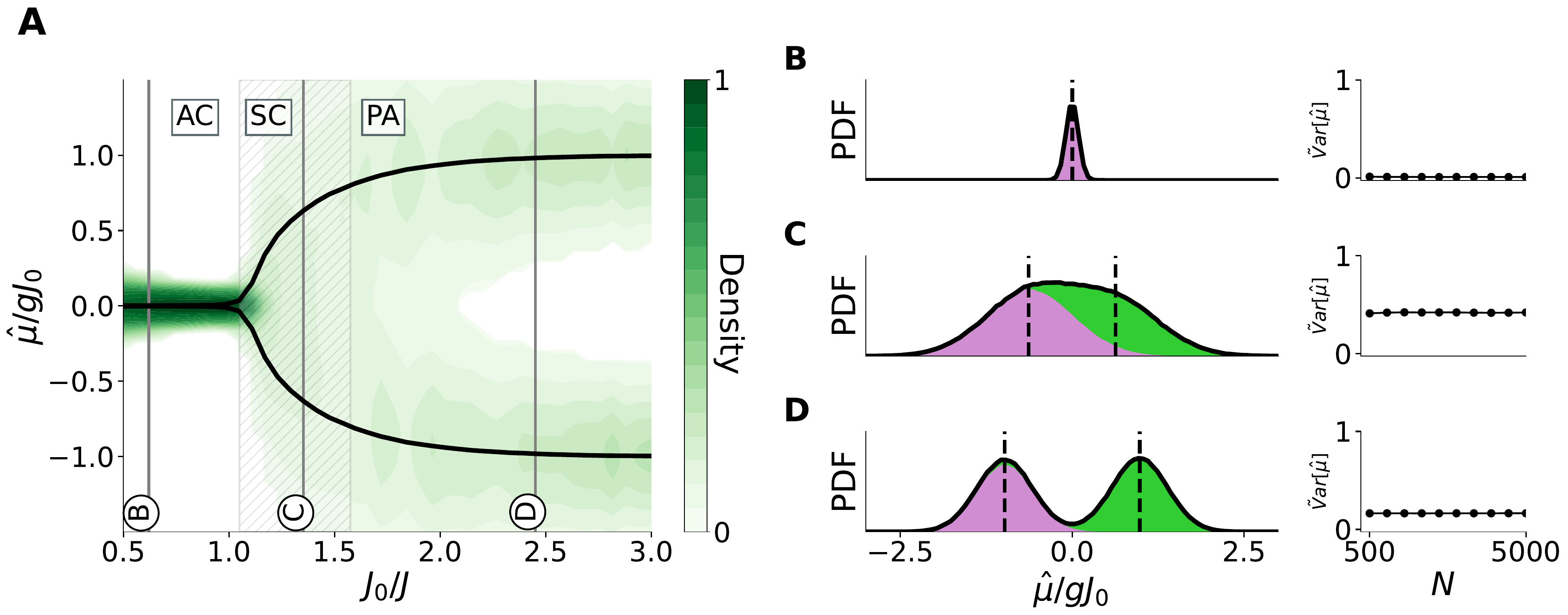}
\caption{Symmetry breaking across dynamical phases. 
\textbf{Panel A}: Heat map of the empirical density of $\hat{\mu}_i/(gJ_0)$ versus $J_0/J$, with black lines showing $\pm \hat{M}$ and the shaded region marking the SC phase. The distribution evolves from unimodal in AC to bimodal in PA, with intermediate broadening in SC.
\textbf{Panels B–D}: Representative histograms at $J_0/J = 0.60, 1.35,$ and $2.45$ (AC, SC, PA). Violet histograms show the Gaussian distribution of the sample-dependent mean for a single network $\mathbf{W}$ realization, centered  —in the broken-symmetry phases-- at $-\hat{M}$, while light-green histograms show the distribution across multiple realizations, such that the positive and negative branches overlap. The variance $\mathrm{Var}[\hat{\mu}]$ is shown as a function of $N$, illustrating that in the SC phase the width vanishes in the thermodynamic limit, while in the AC and PA phases the distribution remains broad, reflecting persistent heterogeneity across units. Simulations used $N = 5000$, $S = 200$, and $T = 2000$.}
    \label{fig:5}
\end{figure}

In Fig.~\ref{fig:5}, the empirical mean input is obtained from microscopic simulations of Eq.~(\ref{eq:scs_dynamics}). For each neuron, it is defined as
\begin{equation} \label{Eq. empirical mu}
\hat{\mu}_i = \frac{1}{T} \int_{t_0}^{t_0 + T} \sum_{j=1}^N W_{ij} x_j(s)\, ds,
\end{equation}
which, for $T \rightarrow \infty$, quantifies the long-time average input received through the disordered connectivity. Figure~\ref{fig:5}A displays the heat map of the empirical density of $\hat{\mu}_i/(gJ_0)$ as a function of $J_0/J$ for $1/gJ = 0.25$. The distribution evolves continuously from unimodal in the AC phase to bimodal in the PA phase, exhibiting an intermediate broadening in the SC regime. In the AC phase, $\hat{\mu}_i/(gJ_0)$ is narrowly distributed, centered at zero ($\hat{M}=0$, black dashed line; Fig.~\ref{fig:5}B) and with a width that diminishes as the system size grows (righmost panel). Upon entering the SC phase, the distribution broadens markedly (Fig.~\ref{fig:5}C), and in the PA regime it becomes distinctly bimodal (Fig.~\ref{fig:5}D). As shown in these panels, the variance of $\hat{\mu}_i$ remains finite as $N$ increases, confirming that the observed broadness in the AC and SC phases is not a finite-size artifact but a genuine manifestation of ergodicity breaking.

The qualitative transition from a narrow to a broad distribution coincides with the bifurcation of the mean activity $\hat{M}$ (black solid line in Fig.~\ref{fig:5}), marking the simultaneous onset of two intertwined phenomena: ergodicity breaking and spontaneous breaking of the $\mathbb{Z}_2$ sign symmetry. The AC–SC transition therefore delineates the joint emergence of disorder-induced freezing and symmetry breaking.

\subsection{Chaotic Dynamics Across Microscopic and Macroscopic Scales}

Let us now trace the dynamical implications of ergodicity breaking in the SCS model by examining the chaotic attractors of the dynamics at both levels of description: microscopic and macroscopic. We demonstrate that, after the mean-activity bifurcation, the single chaotic attractor of the asynchronous-chaos phase splits into a continuum of sample-dependent chaotic attractors, revealing the mechanism behind ergodicity breaking.

\begin{figure}[tbh]
    \centering
    \includegraphics[width=1\linewidth]{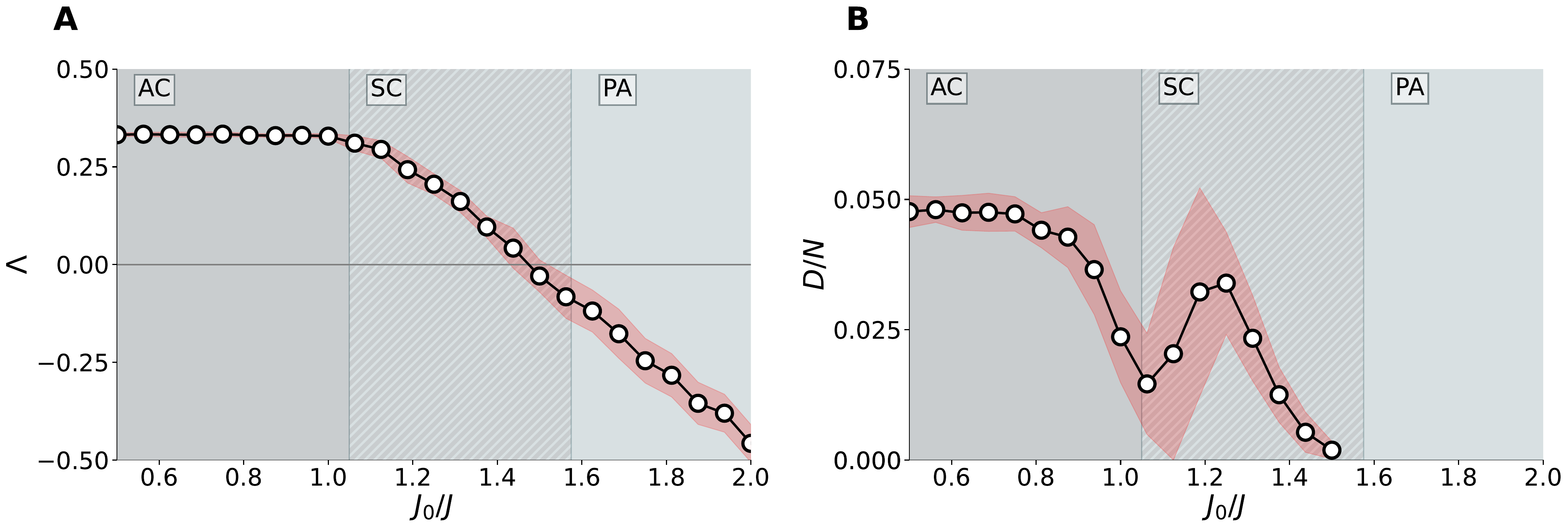}
    \caption{Maximum Lyapunov exponent (\textbf{panel A}) and normalized dimensionality (\textbf{panel B}) as a function of $J_0/J$, computed from the microscopic dynamics, Eq. (\ref{eq:scs_dynamics}), for $1/gJ = 0.25$ and network size $N = 2000$, averaged over $S = 50$ independent realizations. Circles indicate the mean values, while the red shaded region shows variability across realizations at each point. Vertical lines mark the AC–SC and SC–PA transitions, respectively.}
    \label{fig:6}
\end{figure}

 At the microscopic scale, the deterministic high-dimensional dynamics of Eq. (\ref{eq:scs_dynamics}) exhibit classical chaos: single-unit trajectories evolve on a bounded chaotic (strange) attractor with at least one positive Lyapunov exponent \cite{eckmann_ergodic_1985}. Numerical studies confirm that the chaotic regime within the AC phase expands with system size \cite{sompolinsky_chaos_1988, schuecker_optimal_2018, martorell_dynamically_2024}, eventually spanning the entire phase in the thermodynamic limit (see Fig.  \ref{fig:6}A) \footnote{The maximal Lyapunov exponent was computed using the standard Benettin–Wolf algorithm \cite{benettin_lyapunov_1980, wolf_determining_1985, pikovsky_lyapunov_2016}.}. Moreover, chaos is extensive: its dimension scales linearly with  $N$, meaning that as the network grows, there are proportionally more independent directions along which the system exhibits chaotic behavior, while the largest Lyapunov exponents remain independent of system size \cite{engelken_lyapunov_2023, clark_dimension_2023}. In contrast, direct numerical computations or theoretical analyses of Lyapunov spectra in the SC phase have not yet been performed. To address this, our numerical simulations (Fig.  \ref{fig:6}A) confirm the presence of a positive maximal Lyapunov exponent. Notably, the largest exponent exhibits substantial variability across realizations in the SC phase (red shaded region in Fig.  \ref{fig:6}A), reflecting the onset of ergodicity breaking. 
Finally, the Lyapunov exponent becomes negative in the PA phase, consistent with non-chaotic, fixed-point dynamics.

As we have seen, in the large-$N$ limit the collective network dynamics are exactly captured by a self-consistent stochastic DMF equation. Within the chaotic phases, the self-generated Gaussian input $\eta(t)$ exhibits nontrivial temporal correlations, preventing the dynamics from settling into a fixed point. At the macroscopic level (see \ref{App. stability correlation}), linearizing the stationary solution of the DMF equation reveals a single positive Lyapunov exponent: two copies subjected to the same noise but starting infinitesimally apart diverge exponentially. In other words, the combination of (i) persistent stochastic drive and (ii) path-wise sensitivity to initial conditions defines a stochastic chaotic state, the random-dynamical analogue of deterministic chaos \cite{eckmann_ergodic_1985}.
\footnote{A rigorous framework for this notion is provided by the theory of random dynamical systems, where a random attractor combined with a positive sample Lyapunov exponent constitutes stochastic chaos \cite{arnold_random_1998,crauel_attractors_1994}.}
Thus chaos survives the infinite network size limit in the AC phase, consistently with the previously reported results.

The essential difference between the AC and SC chaotic regimes is whether ergodicity is preserved. As already discussed in the AC phase, time and disorder averages coincide, so every realisation explores 
a statistically equivalent chaotic state. However,  once the mean activity bifurcates (\(M\neq0\)), this equivalence is lost: long-time averages retain a static, realisation-dependent component.  The next paragraphs show that this ergodicity breaking translates into a multiplicity of chaotic states at both micro- and macro-scales.

Ergodicity breaking in the SC phase is shown by the static--dynamic decomposition of the effective input, Eq. (\ref{eq:z_split_main}): the fast term \(u(t)\) drives the chaotic fluctuations, whereas the static field \(\zeta\) induces a realisation--dependent bias.  For each fixed \(\zeta\) the DMF dynamics converge---depending on the initial condition \(x(0)\)---to one of
the two symmetric chaotic states that appear after the bifurcation \(\pm M\!\neq\!0\).  Consequently the sample--dependent mean \(\mu(\zeta)\), Eq. (\ref{eq. mu}), takes two possible values depending on $\pm M$. 
Formally, the chaotic attractor in the SC phase is characterized by the collection of disorder-dependent means ${\mu_i(\zeta)}_{i \in \mathbb{N}}$, each corresponding to a unit in the thermodynamic limit. Since these means take non-trivial values when $M \neq 0$ and $\Delta > 0$, the system supports a multiplicity of distinct chaotic attractors across disorder realizations, reflecting the breakdown of ergodicity. By contrast, in the AC phase ($M = 0, \Delta = 0$), all $\mu_i(\zeta)$ collapse to zero, leaving a unique, statistically equivalent attractor that is independent of disorder.

The same structure is observed in finite networks, although with finite-size corrections.  For a given synaptic matrix, the long--time averaged input $\hat{\mu}_i$, Eq. (\ref{Eq. empirical mu}), labels the chaotic state, and its distribution across disorder realizations has width controlled by \(\hat{\Delta}\), Eq. (\ref{Eq. empirical Delta}).  As \(N,T \!\to\! \infty\) the empirical width \(\hat{\Delta}\) converges to the DMF prediction \(\Delta\) (as emphasized by the simulation-theory agreement in Fig. \ref{fig:4}). For finite networks the residual discrepancy can be traced to two finite-size effects: (i) symmetric switching, i.e. rare transitions between the two chaotic branches (\(M\leftrightarrow -M\)) whose mean waiting time grows with \(N\), thereby biasing \(\hat{\mu}\) and slightly inflating \(\hat{\Delta}\); and (ii) metastable coexistence, whereby fluctuations transiently stabilize oscillatory or quasi-fixed states that broaden the \(\hat{\mu}\) distribution, though their lifetimes decrease rapidly with network size and vanish in the thermodynamic limit. Both effects vanish as $N,T \to \infty$, confirming that the multiplicity of attractors is a genuine thermodynamic feature of the SC phase.

Finally, to quantify the effective number of “relevant” dimensions contributing to network variability, we computed the normalized dimensionality using the inverse participation ratio (IPR) of the covariance matrix \cite{hu_spectrum_2022, calvo_frequency-dependent_2024-1, calvo_robust_2025}, defined as
\begin{equation}
D = \frac{\left( \sum_k \lambda_k \right)^2}{\sum_k \lambda_k^2},
\end{equation}
where $\{\lambda_k\}_{k=1}^N$ are the eigenvalues of the zero-lag trajectory-covariance matrix
\begin{equation}
     c_{ij} = \frac 1T \int_{t_0}^{t_0 + T} x_i(s) \, x_j(s) \,  ds  - \hat{m}_i \, \hat{m}_j.
\end{equation}
Intuitively, $D$ measures how many dimensions carry significant variance, providing a compact indicator of the network’s effective degrees of freedom. 

Fig. \ref{fig:6} B displays the normalized dimensionality $D/N$ as a function of $J_0/J$, computed for a microscopic system of size $N = 2000$ driven by Eq. (\ref{eq:scs_dynamics}), and averaged over $S = 50$ independent realizations. In the AC phase, the dimensionality remains roughly constant, consistent with DMF predictions \cite{clark_dimension_2023}. By contrast, in the SC phase, we find that the dimensionality exhibits substantial variability,  reflecting the coexistence of multiple dynamical attractors and the breaking of ergodicity. In the PA regime, deterministic convergence to fixed points produces a singular covariance matrix, rendering the dimensionality ill-defined; this region was therefore excluded from the analysis.

\section{Conclusions and Discussion}
\label{sec: discussion}
The neural model introduced by Sompolinsky, Crisanti, and Sommers (SCS) \cite{sompolinsky_chaos_1988} provides a foundational framework for analyzing the emergence of chaotic dynamics in large random recurrent networks. In its original formulation, interactions strengths are drawn from a Gaussian ensemble with zero mean, $J_0 = 0$, and variance $J$, yielding either a quiescent (Q) or an asynchronous chaotic (AC) phase. Since that seminal work, numerous extensions have been explored, including the introduction of stochastic noise \cite{schuecker_optimal_2018}, the incorporation of structured connectivity such as excitatory–inhibitory populations \cite{harish_asynchronous_2015, kadmon_transition_2015, mastrogiuseppe_intrinsically-generated_2017, mastrogiuseppe_linking_2018}, and the modification of transfer functions \cite{pazo_discontinuous_2024}, among others.

In the present work, we adopt the extension developed in \cite{martorell_dynamically_2024}, where couplings are incorporated as the argument of the nonlinear gain function $\phi(x) = \tanh(x)$, and the mean coupling strength $J_0$ is allowed to be non-zero. This modification introduces a persistent symmetry-breaking component in the effective input  which as the origin of a significantly enriched phase diagram that the model exhibits. In particular, while the standard model can only settle into either a quiescent state, where all units are pinned at zero activity, or a regime of persistent fluctuations called asynchronous chaos, the extended model exhibits two additional phases: one with fixed but heterogeneous non-vanishing activity (PA), and a regime of synchronous chaos (SC), where ongoing fluctuations occur around a non-zero mean, as illustrated in Fig. \ref{fig:1}, which confirms excellent agreement between computational and analytical results. An equivalent SC state, termed ferromagnetic chaos, was recently reported in a related neural model \cite{mastrogiuseppe_linking_2018, fournier_non-reciprocal_2025}, and analogous symmetry-broken chaotic dynamics have also been observed in sparse implementations of the SCS model \cite{metz_dynamical_2025}. These studies highlight that the SC phase is not restricted to the fully connected setting, but rather represents a robust dynamical phenomenon that persists under structural heterogeneity. Our work builds on and extends these insights, providing a unified framework that situates the SC phase within a broader phase diagram and clarifies its connection to spin-glass physics.

Analytically, the collective dynamics are exactly tractable in the thermodynamic limit via a self-consistent dynamical mean-field (DMF) theory. A first analysis, assuming constant two-time autocorrelations, reveals that the associated self-consistent equations --governing the mean activity $M$ and equal-time autocorrelation $q$-- coincide precisely with those of the Sherrington–Kirkpatrick (SK) spin-glass model under the replica-symmetric ansatz \cite{sherrington_solvable_1975}, corresponding to the magnetization and the spin-glass order parameter. As a result, the quiescent phase ($M = 0$, $q = 0$) and its stability boundaries --separating the AC and PA phases-- match exactly the disordered (paramagnetic) phase of the SK model.

The PA phase ($M \neq 0$, $q > 0$ but static) corresponds exactly to the ferromagnetic phase of the SK model, and the transition from PA to SC occurs precisely at the same point as the de Almeida–Thouless line, which in SK marks the onset of the mixed phase where glassy behavior coexists with nonzero magnetization \cite{almeida_stability_1978, mastrogiuseppe_linking_2018}.
The AC phase ($M = 0, \ q > 0$) is analogous to the spin-glass phase of SK: activity fluctuates chaotically around zero, and although the equal-time autocorrelation $C(0)$ is finite, the long-lag autocorrelation decays to zero, reflecting a single, disorder-independent chaotic attractor and requiring a full time-dependent analysis of $C(\tau)$. Between these regimes lies the SC phase, where activity fluctuates around a nonzero mean and the long-lag autocorrelation converges to a finite plateau,
$ C(\tau) \xrightarrow{\tau \longrightarrow \infty} M^2 + \Delta $,
with $\Delta > 0$ capturing residual correlations induced by quenched disorder. This indicates ergodicity breaking in the thermodynamic limit: long-time averages of both the effective input $z_i(t)$ and neuronal activity $x_i(t)$ become neuron-specific. Each unit settles into an idiosyncratic mean, its autocorrelations decay toward that mean squared, and it effectively inhabits its own chaotic trajectory shaped by disorder. Although single trajectories remain internally ergodic, the ensemble across disorder realizations fails to self-average, giving rise to a multiplicity of statistically distinct chaotic attractors. This breakdown of ergodicity at the population level is quantified by $\Delta$, the dynamical analogue of the Edwards–Anderson order parameter, which characterizes frozen disorder in a fully time-dependent setting. These features show that SC is not a trivial extension of the AC phase or of fixed-point dynamics, but a distinct phase with its own organizing principles. The combined extent of the AC and SC phases closely matches the spin-glass and mixed regions in the SK model, with the only difference being that while the boundary between these phases is strictly vertical at $J_0/J = 1$ in SK \cite{nishimori_statistical_2001}, it is very close --but not exactly vertical-- in the SCS model.

The striking similarities and near-perfect mapping between the SCS random recurrent network and the SK spin-glass model are remarkable, given that the SCS model is a (i) dynamical, non-equilibrium system with (ii) continuous variables and (iii) asymmetric couplings, whereas the SK model is a (i) static, equilibrium system with (ii) discrete variables and (iii) symmetric interactions. Despite these profound microscopic differences, the shape of the phase diagram as a function of the control parameters $J_0$ and $J$ is nearly identical for both systems.

A word of caution is warranted when comparing the two systems. In the SK model, the non-ergodic phases are the spin-glass and the mixed phase, while the ferromagnetic phase remains ergodic within each symmetry-broken sector. By contrast, in the SCS model, the AC phase—the counterpart of the SG phase—is ergodic, whereas the PA phase—the counterpart of the ferromagnetic phase—is non-ergodic even within each symmetry-broken sector. Consequently, in the SCS model, the Almeida–Thouless (AT) line does not mark the onset of ergodicity breaking, but rather the emergence of high-dimensional chaos.  Moreover, fundamental differences in the underlying physics are rather obvious: in the SK spin-glass phase, spins remain frozen into many hierarchically nested metastable states, whereas in the analogous phases of the SCS model, the dynamics is inherently chaotic. Thus,  even though the phase diagrams are strikingly similar, the underlying physics of the two systems remains clearly distinct.

These comparisons underscore the novelties of our present work, which builds on the potential-based formulation of the DMF equations—originally introduced by Sompolinsky \cite{sompolinsky_chaos_1988}, that interprets the evolution of the temporal autocorrelation $C(\tau)$ as the motion of a particle in an effective potential $V(C; C_0, M)$. While this approach has traditionally been applied to study fixed-point solutions and their instabilities, we extend it to uncover and characterize the rich dynamics of nontrivial chaotic phases in recurrent networks \cite{crisanti_path_2018, schuecker_optimal_2018, helias_statistical_2020, martorell_dynamically_2024}, we make three main contributions. First, we extend the framework to uncover and characterize SC states beyond the AC phase. Second, we formalize the notion of a dynamically selected state through geometric criteria on the potential, deriving self-consistent conditions on its derivatives (Eqs. (\ref{Eq. V'})-(\ref{Eq. V''})) that uniquely determine the physically admissible steady state. Third, by extending the stability analysis to the SC regime, we show that relaxation toward a non-zero autocorrelation plateau is the only linearly stable solution. Together, these contributions provide a unified geometric framework for the onset and stability of structured chaotic dynamics and clarify the nature of ergodicity breaking in the presence of quenched disorder.

These methods and results contribute to a theoretical understanding of how high-dimensional chaos and ergodicity breaking can arise from quenched disorder in recurrent networks, even in the absence of external input or structural constraints. While the present work focuses on a single homogeneous population with symmetric gain functions, the framework readily extends to more biologically realistic scenarios. Future research could explore the effects of strictly positive, saturating transfer functions, which better approximate neuronal activation, or incorporate excitatory and inhibitory subpopulations with structured connectivity. A key question is whether the synchronous-chaotic phase—and the associated breakdown of ergodicity—persists under these generalizations, and if so, how it shapes the computational and dynamical capabilities of the brain and other networked information-processing systems. Beyond these model-specific extensions, our findings also contribute to the broader debate on the criticality hypothesis in neural systems, by clarifying the spectrum of possible dynamical phases and their transitions \cite{ 
mora_are_2011, plenz_self-organized_2021, munoz_colloquium_2018, chialvo_emergent_2010, shew_functional_2013,  cocchi_criticality_2017, kinouchi_optimal_2006, beggs_cortex_2022, obyrne_how_2022,  morales_quasiuniversal_2023}. Ultimately, these investigations could reveal fundamental principles by which disorder and chaos support robust computation and flexible dynamics, as exploited in reservoir computing and other approaches that harness random recurrent networks for information processing.

\section*{Acknowledgments} 
This work has been supported by Grant No. PID2023-149174NB-I00 financed by the Spanish Ministry and Agencia Estatal de Investigación MICIU/AEI/10.13039/501100011033 and EDRF funds (European Union). We thank Victor Buendía for fruitful discussions. AA acknowledges insightful discussions on the SC phase with Valentina Ros and Pierfrancesco Urbani. 

\section*{References}
\bibliographystyle{iopart-num}
\bibliography{biblioteca}

\providecommand{\newblock}{}
\begin{thebibliography}{10}
\expandafter\ifx\csname url\endcsname\relax
  \def\url#1{{\tt #1}}\fi
\expandafter\ifx\csname urlprefix\endcsname\relax\def\urlprefix{URL }\fi
\providecommand{\eprint}[2][]{\url{#2}}

\bibitem{sompolinsky_chaos_1988}
Sompolinsky H, Crisanti A and Sommers H~J 1988 {\em Physical Review Letters\/}
  {\bf 61} 259--262 ISSN 0031-9007
  \urlprefix\url{https://link.aps.org/doi/10.1103/PhysRevLett.61.259}

\bibitem{sompolinsky_relaxational_1982}
Sompolinsky H and Zippelius A 1982 {\em Physical Review B\/} {\bf 25}
  6860--6875 ISSN 0163-1829
  \urlprefix\url{https://link.aps.org/doi/10.1103/PhysRevB.25.6860}

\bibitem{sompolinsky_dynamic_1981}
Sompolinsky H and Zippelius A 1981 {\em Physical Review Letters\/} {\bf 47}
  359--362 ISSN 0031-9007
  \urlprefix\url{https://link.aps.org/doi/10.1103/PhysRevLett.47.359}

\bibitem{kadmon_efficient_2025}
Kadmon J 2025 {\em Human Arenas\/} ISSN 2522-5790, 2522-5804 publisher:
  Springer Science and Business Media LLC
  \urlprefix\url{https://link.springer.com/10.1007/s42087-025-00507-9}

\bibitem{schuecker_optimal_2018}
Schuecker J, Goedeke S and Helias M 2018 {\em Physical Review X\/} {\bf 8}
  041029 ISSN 2160-3308
  \urlprefix\url{https://link.aps.org/doi/10.1103/PhysRevX.8.041029}

\bibitem{rajan_stimulus-dependent_2010}
Rajan K, Abbott L~F and Sompolinsky H 2010 {\em Physical Review E\/} {\bf 82}
  011903 ISSN 1539-3755, 1550-2376
  \urlprefix\url{https://link.aps.org/doi/10.1103/PhysRevE.82.011903}

\bibitem{pazo_discontinuous_2024}
Pazó D 2024 {\em Physical Review E\/} {\bf 110} 014201 ISSN 2470-0045,
  2470-0053
  \urlprefix\url{https://link.aps.org/doi/10.1103/PhysRevE.110.014201}

\bibitem{van_vreeswijk_chaos_1996}
van Vreeswijk C and Sompolinsky H 1996 {\em Science\/} {\bf 274} 1724--1726
  ISSN 0036-8075, 1095-9203
  \urlprefix\url{https://www.science.org/doi/10.1126/science.274.5293.1724}

\bibitem{harish_asynchronous_2015}
Harish O and Hansel D 2015 {\em PLOS Computational Biology\/} {\bf 11} e1004266
  ISSN 1553-7358
  \urlprefix\url{https://dx.plos.org/10.1371/journal.pcbi.1004266}

\bibitem{kadmon_transition_2015}
Kadmon J and Sompolinsky H 2015 {\em Physical Review X\/} {\bf 5} 041030 ISSN
  2160-3308 arXiv:1508.06486 [cond-mat, physics:nlin, q-bio]
  \urlprefix\url{http://arxiv.org/abs/1508.06486}

\bibitem{mastrogiuseppe_intrinsically-generated_2017}
Mastrogiuseppe F and Ostojic S 2017 {\em PLOS Computational Biology\/} {\bf 13}
  e1005498 ISSN 1553-7358
  \urlprefix\url{https://dx.plos.org/10.1371/journal.pcbi.1005498}

\bibitem{mastrogiuseppe_linking_2018}
Mastrogiuseppe F and Ostojic S 2018 {\em Neuron\/} {\bf 99} 609--623.e29 ISSN
  08966273
  \urlprefix\url{https://linkinghub.elsevier.com/retrieve/pii/S0896627318305439}

\bibitem{stern_reservoir_2023}
Stern M, Istrate N and Mazzucato L 2023 {\em eLife\/} {\bf 12} e86552 ISSN
  2050-084X \urlprefix\url{https://elifesciences.org/articles/86552}

\bibitem{wardak_extended_2022}
Wardak A and Gong P 2022 {\em Physical Review Letters\/} {\bf 129} 048103 ISSN
  0031-9007, 1079-7114
  \urlprefix\url{https://link.aps.org/doi/10.1103/PhysRevLett.129.048103}

\bibitem{martorell_dynamically_2024}
Martorell C, Calvo R, Annibale A and Muñoz M~A 2024 {\em Chaos, Solitons \&
  Fractals\/} {\bf 182} 114809 ISSN 09600779
  \urlprefix\url{https://linkinghub.elsevier.com/retrieve/pii/S0960077924003618}

\bibitem{dick_linking_2024}
Dick M, van Meegen A and Helias M 2024 Linking {Network} and {Neuron}-level
  {Correlations} by {Renormalized} {Field} {Theory} arXiv:2309.14973 [cond-mat]
  \urlprefix\url{http://arxiv.org/abs/2309.14973}

\bibitem{metz_dynamical_2025}
Metz F~L 2025 {\em Physical Review Letters\/} {\bf 134} 037401 ISSN 0031-9007,
  1079-7114
  \urlprefix\url{https://link.aps.org/doi/10.1103/PhysRevLett.134.037401}

\bibitem{fournier_non-reciprocal_2025}
Fournier S~J, Pacco A, Ros V and Urbani P 2025 Non-reciprocal interactions and
  high-dimensional chaos: comparing dynamics and statistics of equilibria in a
  solvable model arXiv:2503.20908 [cond-mat]
  \urlprefix\url{http://arxiv.org/abs/2503.20908}

\bibitem{sherrington_solvable_1975}
Sherrington D and Kirkpatrick S 1975 {\em Physical Review Letters\/} {\bf 35}
  1792--1796 ISSN 0031-9007
  \urlprefix\url{https://link.aps.org/doi/10.1103/PhysRevLett.35.1792}

\bibitem{nishimori_statistical_2001}
Nishimori H 2001 {\em Statistical {Physics} of {Spin} {Glasses} and
  {Information} {Processing}: {An} {Introduction}\/} 1st ed (Oxford University
  PressOxford) ISBN 978-0-19-850941-7 978-0-19-170908-1
  \urlprefix\url{https://academic.oup.com/book/5185}

\bibitem{martorell_supplemental_2025}
 Additional analysis can be found in the Supplemental Material.

\bibitem{martin_statistical_1973}
Martin P~C, Siggia E~D and Rose H~A 1973 {\em Physical Review A\/} {\bf 8}
  423--437 ISSN 0556-2791
  \urlprefix\url{https://link.aps.org/doi/10.1103/PhysRevA.8.423}

\bibitem{de_dominicis_field-theory_1978}
De~Dominicis C and Peliti L 1978 {\em Physical Review B\/} {\bf 18} 353--376
  ISSN 0163-1829
  \urlprefix\url{https://link.aps.org/doi/10.1103/PhysRevB.18.353}

\bibitem{coolen_ch_2001}
Coolen A~C~C 2001 Ch. 15, {Stastistical} {Mechanics} of {Recurrent} {Neural}
  {Networks} {II} - {Dynamics} {\em Neuro-informatics and neural modelling\/}
  ({\em Handbook of biological physics\/} no v. 4) (Amsterdam New York:
  Elsevier) ISBN 978-0-444-50284-1

\bibitem{hatchett_asymmetrically_2004}
Hatchett J~P~L and Coolen A~C~C 2004 {\em Journal of Physics A: Mathematical
  and General\/} {\bf 37} 7199--7212 ISSN 0305-4470, 1361-6447
  \urlprefix\url{https://iopscience.iop.org/article/10.1088/0305-4470/37/29/003}

\bibitem{crisanti_path_2018}
Crisanti A and Sompolinsky H 2018 {\em Physical Review E\/} {\bf 98} 062120
  ISSN 2470-0045, 2470-0053 arXiv:1809.06042 [cond-mat]
  \urlprefix\url{http://arxiv.org/abs/1809.06042}

\bibitem{helias_statistical_2020}
Helias M and Dahmen D 2020 {\em Statistical {Field} {Theory} for {Neural}
  {Networks}\/} ({\em Lecture {Notes} in {Physics}\/} vol 970) (Cham: Springer
  International Publishing) ISBN 978-3-030-46443-1 978-3-030-46444-8
  \urlprefix\url{http://link.springer.com/10.1007/978-3-030-46444-8}

\bibitem{hertz_path_2017}
Hertz J~A, Roudi Y and Sollich P 2017 {\em Journal of Physics A: Mathematical
  and Theoretical\/} {\bf 50} 033001 ISSN 1751-8113, 1751-8121
  \urlprefix\url{https://iopscience.iop.org/article/10.1088/1751-8121/50/3/033001}

\bibitem{zou_introduction_2024}
Zou W and Huang H 2024 {\em SciPost Physics Lecture Notes\/}  79 ISSN 2590-1990
  arXiv:2305.08459 [cond-mat, q-bio]
  \urlprefix\url{http://arxiv.org/abs/2305.08459}

\bibitem{galla_generating-functional_2024}
Galla T 2024 Generating-functional analysis of random {Lotka}-{Volterra}
  systems: {A} step-by-step guide arXiv:2405.14289 [cond-mat, q-bio]
  \urlprefix\url{http://arxiv.org/abs/2405.14289}

\bibitem{edwards_theory_1975}
Edwards S~F and Anderson P~W 1975 {\em Journal of Physics F: Metal Physics\/}
  {\bf 5} 965 ISSN 0305-4608
  \urlprefix\url{https://dx.doi.org/10.1088/0305-4608/5/5/017}

\bibitem{dahmen_second_2019}
Dahmen D, Grün S, Diesmann M and Helias M 2019 {\em Proceedings of the
  National Academy of Sciences\/} {\bf 116} 13051--13060 ISSN 0027-8424,
  1091-6490 number: 26
  \urlprefix\url{http://www.pnas.org/lookup/doi/10.1073/pnas.1818972116}

\bibitem{li_tuning_2020}
Li J and Shew W~L 2020 {\em PLOS Computational Biology\/} {\bf 16} e1008268
  ISSN 1553-7358 number: 9
  \urlprefix\url{https://dx.plos.org/10.1371/journal.pcbi.1008268}

\bibitem{renart_asynchronous_2010}
Renart A, De~La~Rocha J, Bartho P, Hollender L, Parga N, Reyes A and Harris K~D
  2010 {\em Science\/} {\bf 327} 587--590 ISSN 0036-8075, 1095-9203
  \urlprefix\url{https://www.science.org/doi/10.1126/science.1179850}

\bibitem{mieghem_graph_2010}
Mieghem P~V 2010 {\em Graph {Spectra} for {Complex} {Networks}\/} 1st ed
  (Cambridge University Press) ISBN 978-0-511-92168-1 978-0-521-19458-7
  978-1-107-41147-0
  \urlprefix\url{https://www.cambridge.org/core/product/identifier/9780511921681/type/book}

\bibitem{tao_topics_2012}
Tao T 2012 {\em Topics in random matrix theory\/} ({\em Graduate studies in
  mathematics\/} no v. 132) (Providence, R.I: American Mathematical Society)
  ISBN 978-0-8218-7430-1

\bibitem{tao_outliers_2013}
Tao T 2013 {\em Probability Theory and Related Fields\/} {\bf 155} 231--263
  ISSN 0178-8051, 1432-2064
  \urlprefix\url{https://link.springer.com/10.1007/s00440-011-0397-9}

\bibitem{almeida_stability_1978}
Almeida J~R~L~d and Thouless D~J 1978 {\em Journal of Physics A: Mathematical
  and General\/} {\bf 11} 983 ISSN 0305-4470
  \urlprefix\url{https://dx.doi.org/10.1088/0305-4470/11/5/028}

\bibitem{crisanti_dynamics_1988}
Crisanti A and Sompolinsky H 1988 {\em Physical Review A\/} {\bf 37} 4865--4874
  publisher: American Physical Society
  \urlprefix\url{https://link.aps.org/doi/10.1103/PhysRevA.37.4865}

\bibitem{crisanti_dynamics_1987}
Crisanti A and Sompolinsky H 1987 {\em Physical Review A\/} {\bf 36} 4922--4939
  publisher: American Physical Society
  \urlprefix\url{https://link.aps.org/doi/10.1103/PhysRevA.36.4922}

\bibitem{gardiner_handbook_2004}
Gardiner C~W 2004 {\em Handbook of stochastic methods for physics, chemistry,
  and the natural sciences\/} 3rd ed Springer series in synergetics (Berlin New
  York: Springer-Verlag) ISBN 978-3-540-20882-2

\bibitem{borovkov_probability_2013}
Borovkov A~A 2013 {\em Probability {Theory}\/} Universitext (London: Springer
  London) ISBN 978-1-4471-5200-2 978-1-4471-5201-9
  \urlprefix\url{https://link.springer.com/10.1007/978-1-4471-5201-9}

\bibitem{eckmann_ergodic_1985}
Eckmann J~P and Ruelle D 1985 {\em Reviews of Modern Physics\/} {\bf 57}
  617--656 ISSN 0034-6861
  \urlprefix\url{https://link.aps.org/doi/10.1103/RevModPhys.57.617}

\bibitem{benettin_lyapunov_1980}
Benettin G, Galgani L, Giorgilli A and Strelcyn J~M 1980 {\em Meccanica\/} {\bf
  15} 21--30 ISSN 0025-6455, 1572-9648
  \urlprefix\url{http://link.springer.com/10.1007/BF02128237}

\bibitem{wolf_determining_1985}
Wolf A, Swift J~B, Swinney H~L and Vastano J~A 1985 {\em Physica D: Nonlinear
  Phenomena\/} {\bf 16} 285--317 ISSN 01672789
  \urlprefix\url{https://linkinghub.elsevier.com/retrieve/pii/0167278985900119}

\bibitem{pikovsky_lyapunov_2016}
Pikovsky A and Politi A 2016 {\em Lyapunov exponents: a tool to explore complex
  dynamics\/} (Cambridge: Cambridge University Press) ISBN 978-1-107-03042-8

\bibitem{engelken_lyapunov_2023}
Engelken R, Wolf F and Abbott L~F 2023 {\em Physical Review Research\/} {\bf 5}
  043044 ISSN 2643-1564
  \urlprefix\url{https://link.aps.org/doi/10.1103/PhysRevResearch.5.043044}

\bibitem{clark_dimension_2023}
Clark D~G, Abbott L and Litwin-Kumar A 2023 {\em Physical Review Letters\/}
  {\bf 131} 118401 ISSN 0031-9007, 1079-7114
  \urlprefix\url{https://link.aps.org/doi/10.1103/PhysRevLett.131.118401}

\bibitem{arnold_random_1998}
Arnold L 1998 {\em Random {Dynamical} {Systems}\/} Springer {Monographs} in
  {Mathematics} (Berlin, Heidelberg: Springer Berlin Heidelberg) ISBN
  978-3-642-08355-6 978-3-662-12878-7
  \urlprefix\url{http://link.springer.com/10.1007/978-3-662-12878-7}

\bibitem{crauel_attractors_1994}
Crauel H and Flandoli F 1994 {\em Probability Theory and Related Fields\/} {\bf
  100} 365--393 ISSN 0178-8051, 1432-2064
  \urlprefix\url{https://link.springer.com/10.1007/BF01193705}

\bibitem{hu_spectrum_2022}
Hu Y and Sompolinsky H 2022 {\em PLOS Computational Biology\/} {\bf 18}
  e1010327 ISSN 1553-7358
  \urlprefix\url{https://dx.plos.org/10.1371/journal.pcbi.1010327}

\bibitem{calvo_frequency-dependent_2024-1}
Calvo R, Martorell C, Morales G~B, Di~Santo S and Muñoz M~A 2024 {\em Physical
  Review Letters\/} {\bf 133} 208401 ISSN 0031-9007, 1079-7114
  \urlprefix\url{https://link.aps.org/doi/10.1103/PhysRevLett.133.208401}

\bibitem{calvo_robust_2025}
Calvo R, Martorell C, Roig A and Muñoz M~A 2025 Robust {Scaling} in {Human}
  {Brain} {Dynamics} {Despite} {Latent} {Variables} and {Limited} {Sampling}
  {Distortions} arXiv:2506.03640 [q-bio]
  \urlprefix\url{http://arxiv.org/abs/2506.03640}

\bibitem{mora_are_2011}
Mora T and Bialek W 2011 {\em Journal of Statistical Physics\/} {\bf 144}
  268--302 ISSN 1572-9613
  \urlprefix\url{https://doi.org/10.1007/s10955-011-0229-4}

\bibitem{plenz_self-organized_2021}
Plenz D, Ribeiro T~L, Miller S~R, Kells P~A, Vakili A and Capek E~L 2021 {\em
  Frontiers in Physics\/} {\bf 9} ISSN 2296-424X
  \urlprefix\url{https://www.frontiersin.org/articles/10.3389/fphy.2021.639389}

\bibitem{munoz_colloquium_2018}
Muñoz M~A 2018 {\em Reviews of Modern Physics\/} {\bf 90}
  \urlprefix\url{https://link.aps.org/doi/10.1103/RevModPhys.90.031001}

\bibitem{chialvo_emergent_2010}
Chialvo D~R 2010 {\em Nature Physics\/} {\bf 6} 744--750 ISSN 1745-2481 number:
  10 Publisher: Nature Publishing Group
  \urlprefix\url{https://www.nature.com/articles/nphys1803}

\bibitem{shew_functional_2013}
Shew W~L and Plenz D 2013 {\em The Neuroscientist\/} {\bf 19} 88--100 ISSN
  1073-8584 publisher: SAGE Publications Inc STM
  \urlprefix\url{https://doi.org/10.1177/1073858412445487}

\bibitem{cocchi_criticality_2017}
Cocchi L, Gollo L~L, Zalesky A and Breakspear M 2017 {\em Progress in
  Neurobiology\/} {\bf 158} 132--152 ISSN 1873-5118

\bibitem{kinouchi_optimal_2006}
Kinouchi O and Copelli M 2006 {\em Nature Physics\/} {\bf 2} 348--351 ISSN
  1745-2481 publisher: Nature Publishing Group
  \urlprefix\url{https://www.nature.com/articles/nphys289}

\bibitem{beggs_cortex_2022}
Beggs J~M 2022 {\em The {Cortex} and the {Critical} {Point}: {Understanding}
  the {Power} of {Emergence}\/} (The MIT Press) ISBN 978-0-262-37034-9
  \urlprefix\url{https://direct.mit.edu/books/oa-monograph/5372/The-Cortex-and-the-Critical-PointUnderstanding-the}

\bibitem{obyrne_how_2022}
O'Byrne J and Jerbi K 2022 {\em Trends in Neurosciences\/} {\bf 45} 820--837
  ISSN 1878-108X

\bibitem{morales_quasiuniversal_2023}
Morales G~B, di~Santo S and Muñoz M~A 2023 {\em Proceedings of the National
  Academy of Sciences\/} {\bf 120} e2208998120 publisher: Proceedings of the
  National Academy of Sciences
  \urlprefix\url{https://www.pnas.org/doi/10.1073/pnas.2208998120}

\bibitem{price_useful_1958}
Price R 1958 {\em IRE Transactions on Information Theory\/} {\bf 4} 69--72 ISSN
  2168-2712
  \urlprefix\url{https://ieeexplore.ieee.org/abstract/document/1057444}

\bibitem{voigtlaender_general_2020}
Voigtlaender F 2020 A general version of {Price}'s theorem arXiv:1710.03576
  [math] \urlprefix\url{http://arxiv.org/abs/1710.03576}

\end{thebibliography}

\appendix
\newpage
\section*{APPENDICES}
\section{Evaluation of the nonlinear autocorrelation function $\Xi$} \label{App: Gaussian properties}

\subsection{Effective input and Gaussian properties} \label{App. Xi}
For a representative neuron the effective input is defined as
\begin{equation}
   z(t) = gJ_0 M(t) + gJ \eta(t),
\end{equation}
where $M(t)$ is the mean activity and $\eta(t)$ is a Gaussian process with zero mean and covariance $\langle \eta(t)\eta(s)\rangle = C(t,s)$. Consequently, $z(t)$ is also Gaussian, with instantaneous mean $gJ_0 M(t)$ and autocovariance 
\begin{equation}
   \langle z(t)z(s)\rangle - \langle z(t)\rangle \langle z(s)\rangle = (gJ)^2 C(t,s).
\end{equation}

For fixed times $t$ and $s$, the pair $(z(t),z(s))$ has a bivariate Gaussian distribution. To facilitate computations, it is convenient to represent this joint distribution in terms of three independent standard normal variables $u,v,w$ \cite{kadmon_transition_2015}:
\begin{align}
   z(t) \equiv \qquad & z(u, w) =  gJ_0 M(t) + gJ\big(\alpha_1(t,s)\,u + \beta(t,s)\,w\big), \\
   z(s) \equiv \qquad & z(v, w) = gJ_0 M(s) + gJ\big(\alpha_2(t,s)\,v + \beta(t,s)\,w\big),
\end{align}
with coefficients
\begin{align}
   \alpha_1(t,s) &= \sqrt{C(t,t) - \vert C(t,s)\vert}, \\
   \alpha_2(t,s) &= \sqrt{C(s,s) - \vert C(t,s) \vert}, \\
   \beta(t,s)    &= \sqrt{\vert C(t,s) \vert}.
\end{align}
Here $u,v,w \sim \mathcal{N}(0,1)$ are independent. This decomposition diagonalises the covariance matrix and allows averages of arbitrary functions $F(z(t),z(s))$ to be written as Gaussian integrals:
\begin{equation} \label{Eq. average F}
   \langle F(z(t),z(s)) \rangle 
   = \iiint_{-\infty}^{+\infty} Du\,Dv\,Dw \; F(z(u,w), z(v,w)),
\end{equation}
where $D\nu = e^{-\nu^2/2} d\nu / \sqrt{2\pi}$ denotes the standard Gaussian measure.

\subsection{Evaluation of $\Xi$}
\label{App. evaluation Xi}

Setting $F(x,y)=\phi(x)\phi(y)$ and assuming stationarity (meaning that $M(t)\equiv M$ and $C(t,s)\equiv C(\tau)$ for $\tau = \vert t - s \vert$), the nonlinear autocorrelation becomes
\begin{equation}\label{Eq. Xi decomposed}
   \Xi(C(\tau);C(0),M) 
   = \iiint_{-\infty}^{+\infty} Du\,Dv\,Dw \;\phi(z(u,w))\,\phi(z(v,w)),
\end{equation}
with
\begin{align}
   z(u,w) &= gJ_0 M + gJ\left(\sqrt{C(0)-  \vert C(\tau) \vert}\,u + \sqrt{\vert C(\tau) \vert}\,w\right), \nonumber\\
   z(v,w) &= gJ_0 M + gJ\left(\sqrt{C(0)- \vert C(\tau) \vert}\,v + \sqrt{ \vert C(\tau) \vert}\,w\right). \label{Eq. x and y}
\end{align}

Thus, the nonlinear autocorrelation $\Xi$ is reduced to a scalar-valued Gaussian integral depending only on the lag $\tau$, the zero-lag variance $C(0)$, and the mean activity $M$. This compact formulation is suitable for both analytical treatment and numerical evaluation.

\subsection{Properties of $\Xi$ and the Price's theorem} 
\label{App: Price th}
Function $\Xi(C; C_0, M)$, as defined by the integral form given in Eq. (\ref{Eq. Xi decomposed}), is a scalar-valued function of variable $C \in \mathbb{R}$, with $C_0 = C(0)$ and $M$ fixed. A crucial constraint on the domain of $C$ arises from the requirement that the covariance matrix of the bivariate Gaussian distribution must remain symmetric and positive semi-definite. This condition imposes the inequality $|C| \leq C_0$ where equality holds at zero lag, $\tau = 0$, ensuring that the zero-lag autocorrelation represents the maximal correlation. As noted, this physical condition bounds the domain of $\Xi$ function. In the following, we list some key properties of $\Xi$ function. 

\subsubsection*{Symmetry}
Because the gain function is odd, \( \phi(-x)=-\phi(x) \), function $\Xi$ is an even function of the mean activity:
\begin{equation}
  \Xi(C;C_{0},-M)=\Xi(C;C_{0},M).
\end{equation}
Moreover, once \(C_{0}^{\text{sel}}\) and \(M\) are fixed self-consistently by Eqs. (\ref{Eq. V'})–(\ref{Eq. V''}) and (\ref{Eq. M}), $\Xi$ satisfies
\begin{equation}
  \Xi\bigl(0;C_{0}^{\text{sel}},M\bigr)=M^{2}.
\end{equation}
In the particular case \(M=0\) the symmetry implies \(\Xi(0;C_{0},0)=0\) for any value of \(C_{0}\).

\subsubsection*{Derivatives representation -- Price's Theorem} 
It is worth-noting the expression derived from the derivatives of $\Xi$ as it involves a useful general expression, known as Price's theorem (published in \cite{price_useful_1958}, see also a modern extension of this result in \cite{voigtlaender_general_2020}). The first derivative at $C$ is given by
\begin{eqnarray}
    \partial_{C} \Xi(C; C_0, M) = & \iiint_{-\infty}^{+\infty} Du Dv Dw  &\big[  \phi'(z(u, w)) \phi(z(v, w)) \, \partial_{C} z(u, w) \nonumber \\
    & &+\phi(z(u, w)) \phi'(z(v, w)) \, \partial_{C} z(v, w) \big] \nonumber 
\end{eqnarray}
where $z(u, w)$ and $z(v, w)$ are defined in Eq. (\ref{Eq. x and y}). Computing $\partial_{C} z(u, w)$ and $\partial_{C} z(v, w)$, it leads to
  
\begin{eqnarray}
\partial_{C} \Xi(C; C_0, M)
    &= \displaystyle-\frac{1}{2} \frac{g J}{\sqrt{C_0 - \vert C \vert}} \iint Dv Dw \, \int du \partial_u(-p(u))\, \phi'(z) \phi(z) - \nonumber\\
    & \displaystyle - \frac{1}{2} \frac{g J}{\sqrt{C_0 - \vert C \vert}} \iint Du Dw \, \int dv \partial_v(-p(v))\, \phi(z) \phi'(z) + \nonumber\\
    & -\frac{1}{2} \frac{g J}{\sqrt{ \vert C \vert}} \iint Du Dv  \int dw \partial_wp(w) (\phi'(z) \phi(z) + \phi(z) \phi'(z))  \nonumber
\end{eqnarray}
where $p(x) = \exp(-x^2/2)/\sqrt{2 \pi}$ and $\phi' (x)$ stands for the first derivative of $\phi(x)$. Therefore, integrating each term by parts, the previous expressions reduces to 
\begin{equation} \label{eq:Xi_first_derivative}
      \partial_{C} \Xi(C; C_0, M)  = (g J)^2 \, \iiint DuDvDw \, \phi'(z(v, w)) \phi'(z(u, w)) 
\end{equation}
By following the same procedure, we can obtain the general expression for higher derivatives --result known as the Price's theorem \cite{voigtlaender_general_2020}:  
\begin{equation}
      \partial^n_{C} \Xi(C; C_0, M)  = (g J)^{(2n)} \,\iiint DuDvDw\, \phi^{(n)}(z(u, w)) \phi^{(n)}(z(v, w)) 
\end{equation}
where $\phi^{(n)}$ stands for the $n$-th derivative of $\phi$. 

Additionally, the derivative of $\Xi$ with respect to the parameter $C_0$ is evaluated as
\begin{equation} \label{eq:Xi_C0_derivative}
    \partial_{C_0} \Xi(C; C_0, M) = \frac{(g J)^2}{2} \,\iiint DuDvDw\,  \phi''(z(u, w)) \phi(z(v, w)) + \phi(z(u, w)) \phi''(z(v, w))
\end{equation} 

These derivatives are particularly useful in determining the properties and shape of the potential $V$, defined by the self-consistency equation (\ref{Eq. descent-gradient expression}). For instance, considering the special case of a fixed-point solution (i.e., a stationary regime in which $C(\tau)=C_0=q$), the derivatives simplify substantially. Specifically, substituting $C=q$ into equations \eqref{eq:Xi_first_derivative}–\eqref{eq:Xi_C0_derivative} yields: 
\begin{eqnarray}
\partial_{C}\Xi(q;q,M)&=&(gJ)^2\int Du\left[\phi'(gJ\sqrt{q}u+gJ_0M)\right]^2, \label{Eq. C' fixed point}\\
\partial_{C}^2\Xi(q;q,M)&=&(gJ)^4\int Du\left[\phi''(gJ\sqrt{q}u+gJ_0M)\right]^2,\\
\partial_{C_0}\Xi(q;q,M)&=&(gJ)^2\int Du \phi''(gJ\sqrt{q}u+gJ_0M)  \phi(gJ\sqrt{q} u+gJ_0 M)    
\end{eqnarray} 
where the integration measure $Du$ corresponds to averaging over the standard Gaussian variable $u$.

\subsubsection*{Monotonicity and upper end-point evaluation}
Function $\Xi(C; C_0, M)$ is monotonically increasing due to Price's theorem, meaning that for $C_1 < C_2$, 
\begin{equation}
    \Xi(C_1; C_0, M) < \Xi(C_2; C_0, M)
\end{equation}
Moreover, for $C_0 \leq q$, the monotonic condition leads to the following inequality, 
\begin{equation} 
    C_0\leq\Xi(C_0; C_0, M) \leq q
\end{equation}
where equality holds only when $C_0 = q$. 

\subsubsection*{Boundedness}
Because of monotonicity, for $C_0 < q$ and $\vert C \vert < C_0$, then 
\begin{equation}
    \vert \Xi(C; C_0, M) \vert < q
\end{equation}

\section{Linear stability analyses}
\label{App:SCS_Jacobian}
The linear stability of any stationary solution $x(t)$ is derived  by analyzing the linear response of the macroscopic observables --mean activity $M$ and autocorrelation $C(0)$ -- to external perturbations. Since those external perturbations may break time-translational symmetry, the following analysis is done considering the general, non-stationary expression. 

\subsection{Linear stability of mean activity $M$} \label{App: jacobian_M}
Starting from the DMF equation (\ref{eq:dmft_eq}), the evolution of the mean activity is given by
\begin{equation}
\partial_t M(t)= -M(t) + \Sigma \bigl(C(t, t), M(t)\bigr),
\label{eq:App_meanEvo}
\end{equation}
with the nonlinear kernel
\begin{equation}
\Sigma \bigl(C(t, t), M(t) \bigr)=\int D\eta\;\phi \bigl(gJ_{0}M(t) +gJ\sqrt{C(t, t)} \, \eta\bigr).
\label{eq:App_Sigma}
\end{equation}

Let's assume the system has reached a stationary regime after a transient time, such that the mean activity $M(t)$ becomes time-independent,  $M$; and the two-time autocorrelation function $C(t,s)$ becomes time-translation invariant, $C(t, s)  \equiv C(t - s)$. By perturbing slightly the mean activity, we can analyze how the system linearly respond to this change. This is done by substituting $M \longrightarrow M + \varepsilon\, \delta M(t)$, being $\varepsilon > 0$ a small parameter quantifying the perturbation strength and $\delta M(t)$ the time-dependent, linear perturbation. When inserted into Eq.\eqref{eq:App_meanEvo} results in: (i) terms of order $\mathcal{O}(\varepsilon^0)$, which yield Eq. (\ref{Eq. M 2}), and (ii) terms of order $\mathcal{O}(\varepsilon)$, leading to the following equation, 
\begin{equation}
\partial_t \, \delta M =\Bigl[\partial_M\Sigma\bigl(C(0),M\bigr)-1\Bigr]\delta M ,
\label{eq:App_deltamean}
\end{equation}
with
\begin{equation}
\partial_M\Sigma\bigl(C(0),M\bigr)= gJ_{0}\int D\eta\;\phi' \bigl(gJ_{0}M+gJ\sqrt{C(0)}\, \eta\bigr).
\label{eq:App_dSigma}
\end{equation}
Equation \eqref{eq:App_deltamean} shows that the perturbation decays ($\delta M\to 0$) if
\begin{equation}
\partial_M\Sigma\bigl(C(0),M\bigr)<1.
\label{eq:App_stabCriterion}
\end{equation}
At the AC phase, where $M = 0$, the stability condition reduces to 
\begin{equation}
    g J_0 \, \int D\eta \; \phi'\left(g J \sqrt{C_0^{\text{sel}}} \, \eta\right) < 1.
\end{equation}
At the AC-SC phase transition, the mean activity bifurcates implying, therefore, condition (\ref{eq:AC_SC_main}).

For fixed–point states, the autocorrelation is given by $C(t,s)\equiv q$, so that Eq.\eqref{eq:App_dSigma} reduces to
\begin{equation}
\partial_M\Sigma(q,M)=gJ_{0}\bigl[1-q\bigr],
\end{equation}
where the identity $\phi'(x)=1-\phi^{2}(x)$ has been used together with the definition of $q$ (Eq.\ref{Eq. q}).
Because $0\le q\le 1$, the stability condition \eqref{eq:App_stabCriterion} becomes
\begin{equation}
gJ_{0}<1,
\end{equation}
which coincides with the criterion obtained in Sec. \ref{Sec. fixed-point analysis}. Thus, in the SCS model a fixed point is linearly stable if, and only if, the effective mean coupling satisfies $gJ_{0}<1$.

\subsection{Linear stability of stationary autocorrelation $C(\tau)$} \label{App. stability correlation}
Following the methodological steps outlined in \cite{martorell_dynamically_2024,schuecker_optimal_2018}, the stability of stationary trajectories is analyzed by considering small perturbations \( \varepsilon \, \delta(t,s)\) that break time-translation invariance. Under stationary conditions, the autocorrelation is described by  
\begin{align}
C(t,s) &= C(\tau) + \varepsilon\,  \delta C(t,s), \label{eq:C_p} \\
C(t,t) &= C(s,s) \equiv C(0), \label{eq:C0_p} \\
M(t) & \equiv  M, \label{eq:M_p}
\end{align}
where \(\tau = t - s\). These expressions represents a non-stationary regime where one-time quantities, e.g. $M(t)$, remain constant while two-time quantities, e.g. $\delta C(t,s)$,  break time-translation invariance.

The equation of motion for such a non-stationary state is described by a constant $M$ and the two-time derivative of $C(t, s)$:
\begin{equation} \label{eq:F}
  (1+\partial_t)(1+\partial_s)C(t,s) =  \langle \phi(z(t)) \phi(z(s)) \rangle 
\end{equation}

Substitution of Eqs. \eqref{eq:C_p}, \eqref{eq:C0_p}, and \eqref{eq:M_p} into Eq. \eqref{eq:F}, followed by Taylor expansion of the right-hand side to linear order in \(\varepsilon\), and the use of the symmetry \( C(\tau) = C(-\tau) \), results in: (i) terms of \({\mathcal O}(\varepsilon^0)\), which yield Eq. \eqref{Eq. evolution Delta}, and (ii) terms of \({\mathcal O}(\varepsilon)\), leading to the following equation -- describing the deviation from stationarity--, 
\begin{equation}
(1+\partial_t)(1+\partial_s)\delta C(t,s) = 
\partial_C \Xi(C(\tau),C_0,M)\, \delta C(t,s).
\label{eq:dot_delta}
\end{equation}
The perturbation $\partial C(t, s)$ therefore obeys a linear partial differential equation with coupled time derivatives. As we will show later, it decays asymptotically provided that
\begin{equation}
    \partial_C \Xi(C(\tau),C_0,M) < 1
\end{equation}
which constitutes the necessary and sufficient criterion for linear stability of the stationary autocorrelation. For fixed point solutions ($C(\tau) \equiv q$), replacing by Eq.(\ref{Eq. C' fixed point}), the stability condition reduces to
\begin{equation}
   (g J)^2\,  \int D\eta \left[\phi' (g J_0 M + g J \sqrt{q}\eta )\right]^2 < 1
\end{equation} 
recovering the bifurcation line Eq. (\ref{eq:PA_SC_main}) at the SC-PA transition. Note also that at the Q-AC transition, since $M = 0$, Taylor expanding $\phi'(x)$ for small values of $q$, the stability criterion leads to
\begin{equation}
    (g J)^2 < 1
\end{equation}
which coincides with the criterion obtained in Sec.\ref{Sec. fixed-point analysis} for fixed-point stability.  

\subsection{Precise derivation of $C(\tau)$ stability -- Lyapunov exponents}
We can introduce the transformations \( T = t + s \) and \( \tau = t - s \), and define the ansatz
\begin{equation}
\delta C(t, s) \equiv \psi(\tau)e^{\kappa T/2},
\end{equation}
where $\kappa$ plays the role of the maximal Lyapunov exponent, as a response to a small perturbation on $T$. Eq. \eqref{eq:dot_delta} yields a Schrödinger-like equation for \(\psi(\tau)\), 
\begin{equation}
[-\partial_\tau^2 - V''(C(\tau); C_0,M)]\psi(\tau) =
\left[1 - \left(\frac{\kappa}{2} + 1\right)^2\right]\psi(\tau),
\end{equation}
with an effective quantum mechanical potential \(-V''(C; C_0,M)\), the second derivative of the potential given by Eq.(\ref{Eq. potential}), i.e. $1 - \partial_C \Xi(C; C_0,M)$.

The allowed eigenvalues correspond to energies \( E_0 < E_1 < E_2 \ldots \), where \( E_n = 1 - (\frac{\kappa_n}{2} + 1)^2 \), with associated eigenfunctions \(\psi_n(\tau)\). Stability of the stationary solution \(C(\tau)\) requires the ground-state energy, denoted by $E_0$, to be positive; or, equivalent, \(\kappa_0 = -1 + \sqrt{1 - E_0}\) to be negative. 

For a well-defined eigenfunction $\psi_n(\tau)$, $E \geq -V''(C(\tau); C_0, M)$ for any $\tau$. Therefore, if $V''(C(\tau); C_0, M) < 0$, it guarantees the stability of such an eigenfunction. 

The following argument used the fact that, by Eq. \eqref{Eq. evolution Delta}, 
    \begin{equation}
        [-\partial_\tau^2 - V''(C(\tau)|C_0,M)] \, \partial_\tau C(\tau) = 0,
        \label{eq:Cdot}
    \end{equation}
indicating that non-fixed-point steady-state solutions ($\partial_\tau C(\tau) \neq 0 $) correspond to eigenfunctions with zero energy --when they exist. Therefore, the dynamical properties of those solutions can be used to characterize the ground-state energy. Two specific cases are analyzed:  
\begin{enumerate}
    \item \textbf{Case \(M = 0\):} 
    The potential $V(C; C_0, M)$ exhibits at least one minimum, either at \(C = 0\) or \(C = \pm \bar{C}(C_0)\), for any \(C_0 \leq q\). 

    On one hand, when $C_0 = q$, $V''(q;q, M) > 0$ implying that the associated energy is negative. Therefore, the ground-state energy, for this choice of $C_0$, is negative. 

    For \(C_0 < q\), the potential allows non-trivial solutions of Eq. \eqref{eq:Cdot} corresponding to periodic orbits or asymptotic motion. Periodic solutions exhibit multiple nodes (as $\partial_\tau C(\tau)$ vanishes at each turning point), while asymptotic trajectories have one node (where $\partial_\tau C(\tau) = 0$). A well-known result on Quantum Mechanics (also referred as Sturm-Liouville theory) states that noiseless solutions with at least one node correspond to excited states (\(n \geq 1\)), implying negative ground-state energy (\(E_0 < 0\)). 

    As a consequence, all physical steady-state solutions are unstable in the noiseless case against small perturbations that break time-translation invariance, leading to the emergence of chaos. That chaotic solution is, indeed, described by the asymptotic motion ($C_0 = C_0^{\text{sel}}$), corresponding to a \textit{unstable} stationary state where any slight perturbation causes it to move away from the separatrix curve, as explained in \cite{martorell_dynamically_2024}. As demonstrated in that article, the system dynamically selects that solution. 

    \item \textbf{Case \(M \neq 0\):} Two scenarios arise:  
    \begin{enumerate}
        \item If \(V'(C, C_0,M)\) is strictly decreasing for \(-C_0 \leq C \leq C_0\), \(V''(C|C_0,M) < 0\), ensuring \(E_n > 0 , \; \forall n\). This implies the only solution of Eq. \eqref{eq:Cdot} is the stable fixed point \(C(\tau) = q\).  
        \item If \(V(C, C_0,M)\) contains a well, the analysis follows the same reasoning as in the \(M = 0\) case, concluding that all physical steady-state solutions are unstable. Hence, the system dynamically selects a \textit{unstable} stationary state that asymptotically converges to the separatrix curve. 
    \end{enumerate}
\end{enumerate}

\subsection{Joint perturbations in the AC phase}
 \label{App: extended Jacobian analysis}
We extend the linear stability analysis to simultaneous perturbations of the mean activity $M(t)$ and the equal-time autocorrelation $Q(t)\equiv C(t,t)$ around the asynchronous-chaotic (AC) state. Related approaches based on linear response functions have been developed in \cite{kadmon_transition_2015,mastrogiuseppe_linking_2018}, leading to equivalent stability criteria. Here we provide an explicit Jacobian formulation to validate the simplified linear stability analysis done before. 

Linearizing Eq. \ref{eq:App_meanEvo} yields
\begin{equation}
\partial_t\,\delta M(t)=
\Big(\partial_M\Sigma(Q,M)-1\Big)\,\delta M(t)
+\Big(\partial_Q\Sigma(Q,M)\Big)\,\delta Q(t),
\end{equation}
with $\partial_M\Sigma=gJ_0\langle \phi'(z)\rangle$ and $
\partial_Q\Sigma=\tfrac{(gJ)^2}{2}\langle \phi''(z)\rangle$ where $z\sim\mathcal N(gJ_0M,(gJ)^2Q)$.

From Eq. \ref{eq:F}, symmetry of $C(t,s)$ and 
\begin{equation*}
\partial_t Q(t)=\lim_{s\to t}(\partial_t+\partial_s)C(t,s)=2\lim_{s\to t}\partial_t C(t,s),
\end{equation*}
one obtains
\begin{equation*}
\partial_t\,\delta Q(t)=
2\big(\partial_C\Xi+\partial_{Q(t)}\Xi+\partial_{Q(s)}\Xi\big)\Big|_{s\uparrow t}\,\delta Q(t)
+2\big(\partial_{M(t)}\Xi+\partial_{M(s)}\Xi\big)\Big|_{s\uparrow t}\,\delta M(t)
-2\,\delta Q(t).
\end{equation*}
where the last term arises from the double leakage of $(1+\partial_t)(1+\partial_s)$ at equal times. Using the Gaussian identities of \ref{App: Gaussian properties},
\begin{equation}
\partial_C\Xi=(gJ)^2\langle(\phi'(z))^2\rangle,\; 
\partial_{Q(s)}\Xi=\tfrac{(gJ)^2}{2}\langle \phi(z)\,\phi''(z)\rangle,\; 
\partial_{M(s)}\Xi=\tfrac{gJ_0}{2}\langle \phi(z)\,\phi'(z)\rangle.
\end{equation}

The coupled dynamics $(\delta M,\delta Q)$ are thus governed by the Jacobian
\begin{equation}
\mathcal{J}=
\begin{pmatrix}
gJ_0\langle \phi'(z)\rangle -1 & \tfrac{(gJ)^2}{2}\,\langle \phi''(z)\rangle \\[6pt]
2gJ_0\langle \phi(z)\,\phi'(z)\rangle & (gJ)^2\big(\langle(\phi'(z))^2\rangle+\langle \phi(z)\,\phi''(z)\rangle\big)-2
\end{pmatrix}.
\end{equation}

In the AC state ($M=0$), the input signal $z(t)$ is centered. Since $\phi = \tanh(x)$ is odd, one has $\langle \phi\rangle=0$ and, by parity, $\langle \phi\,\phi'\rangle=0$. Hence $\mathcal{J}_{QM}=0$ and the Jacobian triangularizes, with eigenvalues
\begin{equation}
\lambda_M=gJ_0\langle \phi'(z)\rangle -1,\qquad
\lambda_Q=(gJ)^2\big(\langle(\phi'(z))^2\rangle+\langle \phi(z)\,\phi''(z)\rangle\big)-2.
\end{equation}
Stability requires $\lambda_M<0$ and $\lambda_Q<0$. The AC$\to$SC transition is triggered by the neutralization of $\lambda_M$, corresponding to spontaneous symmetry breaking of $M$, and yields Eq. (\ref{eq:AC_SC_main}). Including $\delta Q$ thus leaves the bifurcation condition unchanged, while confirming that stability also requires damping of equal-time fluctuations ($\lambda_Q<0$).

\section{Ergodic theory and disorder averages}

\subsection{Ergodicity and Time Averages}
\label{app:ergodicity}

For the dynamical mean-field equation Eq. (\ref{eq:dmft_eq}), the evolution of the effective neuron $x(t)$ consists in a stationary continuous-time Markov process. Here we  present the formal derivations that underlie the discussion in Sec. \ref{subsec:theory_SC_main}, introducing the mathematical notions required to establish its ergodic character.

A stationary continuous‑time Markov process is called ergodic when (i) it is irreducible, meaning that every state can be reached in finite time with non‑zero probability from any initial condition, and (ii) it possesses a unique invariant probability distribution. These conditions ensure that the dynamics revisits the entire state space without becoming trapped, so a sufficiently long trajectory reproduces the statistics of the ensemble \cite{gardiner_handbook_2004,eckmann_ergodic_1985}.

Let $x(t)$ denote a stationary and ergodic Markov process. The long–time average along one trajectory is defined as
\begin{equation}
   \overline{x}_{T}= \frac{1}{T}\int_{0}^{T}x(t)\,dt .
   \label{eq:time_average_def}
\end{equation}
The Birkhoff–Khinchin's theorem guarantees that the long-time average coincides almost surely with the ensemble-average of the process at an arbitrary time $t$ \cite[Sec. 16.3]{borovkov_probability_2013}: 
\begin{equation}
   \lim_{T\to\infty}\overline{x}_{T}= \langle x(t)\rangle
   \qquad\text{a.s.}
   \label{eq:ergodic_theorem}
\end{equation}
Moreover, the long-time variance converges almost surely to 0,
\begin{equation}
\overline{(x^2)}_{T}
      -\langle x(t)\rangle^{2}
     \xrightarrow[T\to\infty]{}0 \qquad\text{a.s.}
     \label{eq:ergodic_variance}
\end{equation}
In practice this implies that empirical averages over a single, sufficiently long trajectory coincide with averages computed over many independent realizations. 
A convenient diagnostic for ergodicity, particularly in Gaussian process, is the absolute integrability of the stationary autocorrelation function of $x(t)$,
\begin{equation}
\int_{0}^{\infty}|C(\tau)| \mathrm d\tau<\infty .
\label{eq:app_int_corr}
\end{equation}
When this condition holds, temporal correlations decay sufficiently fast, which guarantees the mixing property (the system “forgets” its past on a finite correlation timescale), thereby validating Eqs.\eqref{eq:ergodic_theorem}-\eqref{eq:ergodic_variance}. 

\subsection{Disorder-Averaged Statistics}
\label{app:disorder}

Averaging \eqref{eq: inner mean} over the quenched variable
\(\zeta\) yields
\begin{equation}
   \langle m(\zeta)\rangle_{\zeta}
   =\int_{-\infty}^\infty D \eta\,
     \phi\; \bigl(gJ_{0}M+gJ C(0)\; \eta\bigr)
   =M,
\end{equation}
where $D=\exp[-\eta^{2}/2]/\sqrt{2\pi} d \eta$ denotes the normal density \(\mathcal N(0,C(0))\) of the Gaussian process \(\eta(t)\) at a fixed time.

To compute the variance of \(m(\zeta)\), we decompose
\(\eta(t)=u(t)+\zeta\) in the dynamical mean-field kernel
\(\Xi(C)=\langle\phi(z(t))\phi(z(t+\tau))\rangle\).
For large \(\tau\) the terms \(u(t)\) and \(u(t+\tau)\) decorrelate
(\(\tilde C(\tau)\to0\)), whereas the quenched variable \(\zeta\) is fixed at both times, giving
\(\Xi(C_\infty)=\langle m(\zeta)^{2}\rangle_{\zeta}\).
Because \(C_\infty\) satisfies the self-consistent DMF condition
\(C_\infty=\Xi(C_\infty)\),
\begin{equation}
   \Delta
   \;=\;
   C_\infty-M^{2}
   \;=\;
   \operatorname{Var}_{\zeta}[m(\zeta)] .
   \label{app:eq:Delta_variance}
\end{equation}

Using the Birkhoff-Khinchin's result with the previous identity, we obtain
\begin{equation}
   \lim_{T\to\infty}
   \operatorname{Var}_\zeta\!\Bigl[
     \frac{1}{T}\!\int_{0}^{T}\!x(t)\,dt
   \Bigr]
   \;=\;
   \operatorname{Var}_{\zeta}[m(\zeta)]
   \;=\;
   \Delta .
\end{equation}

\subsection{Row–projection derivation of the DMFT input} \label{App: row-projection}

In the following, we consider a finite-size network in the SC phase and fix a single realization of the disorder. Inspired by the simulation-based analysis of Sec. \ref{subsec:theory_SC_main}, the neural activity is decomposed as  
\[
x_j(t) = \hat{m}_j + \tilde{x}_j(t),
\]  
where the neuron-specific inner means satisfy  
\begin{equation}
    \hat{M} = \frac{1}{N} \sum_{j=1}^N \hat{m}_j, 
    \qquad   
    \hat{\Delta} = \frac{1}{N} \sum_{j=1}^N (\hat{m}_j - \hat{M})^2,
\end{equation}  
and the fluctuations obey  
\begin{equation}
    \lim_{T \to \infty} \frac{1}{T}\int_0^T \tilde{x}_i(s)\,ds = 0,
    \qquad 
    \lim_{T \to \infty} \frac{1}{T}\int_0^T \tilde{x}_i(s)\,\tilde{x}_j(s+\tau)\,ds = \delta_{ij}\,\tilde{C}(\tau),
\end{equation}  
with  
\[
\tilde{C}(\tau) = \hat{C}(\tau) - \hat{M}^2 - \hat{\Delta}.
\]  

For Gaussian synaptic couplings
\(W_{ij}=J_0/N + J\,\delta W_{ij}/\sqrt N\), with
\(\delta W_{ij}\stackrel{\text{i.i.d.}}{\sim}\mathcal N(0,1)\), the microscopic
current to neuron \(i\) is
\begin{equation*}
z_i(t) = g\sum_{j}W_{ij}\,x_j(t) = gJ_0 \hat{M} +gJ\,\frac1{\sqrt N}\sum_{j}\delta W_{ij} \hat{m}_j +gJ\,\frac1{\sqrt N}\sum_{j}\delta W_{ij}\tilde{x}_j(t).
\end{equation*}

The first sum is a projection of the Gaussian \(i\)-th weight row onto the static pattern \(\{\hat{m}_j\}\); hence, by the central-limit theorem
\begin{equation}
\hat{\zeta}_i\equiv N^{-1/2}\sum_j\delta W_{ij}m_j
\stackrel{d}{\longrightarrow}\mathcal N(0,M^{2}+\Delta). 
\end{equation}

The second sum defines the zero-mean colored noise
\begin{equation}
\hat{u}_i(t)\equiv N^{-1/2}\sum_j\delta W_{ij}\tilde{x}_j(t)
\end{equation}
with \(\langle u_i(t)u_i(s)\rangle=\tilde C(t-s)\). Thus, in the thermodynamic limit, the signal input converges to the DMF terms. 
\begin{equation}
   z_i(t) \longrightarrow gJ_0 M+gJ\,\zeta +gJ\,u(t),
\end{equation}
where $ \zeta \sim\mathcal N(0,M^{2}+\Delta),\;
   \langle u(t) {u}(s)\rangle= \tilde{C}(t-s).$

The decomposition therefore reproduces exactly the DMFT input
 and clarifies the geometric origin of the static offset as a row-wise projection of quenched connectivity onto the static component of the population activity. In particular, since $i$-th row is independent of $j$-th row, $\zeta_i$ and $\zeta_j$ are uncorrelated. 

\end{document}